\documentstyle[eqsecnum,prd,aps,epsfig]{revtex}
\begin{document}
%
\draft
\title{Boltzmann equations for neutrinos with flavor mixings}
%
\author{Shoichi Yamada 
\thanks{e-mail : shoichi@phys.s.u-tokyo.ac.jp, 
TEL : +81-3-5841-4191, 
FAX : +81-3-5841-4224}
}
\address{Research Center for the Early Universe (RESCEU), 
School of Science, University of Tokyo\\
7-3-1 Hongo, Bunkyo-ku, Tokyo 113-0033, JAPAN}
\date{\today}
\maketitle
\begin{abstract}
With a view of applications to the simulations of supernova explosion
and proto neutron star cooling, we derive the Boltzmann equations for 
the neutrino transport with
the flavor mixing based on the real time formalism of the
nonequilibrium field theory and the gradient expansion of the Green
function. The relativistic kinematics is properly
taken into account. The advection terms are derived in the mean field
approximation for the neutrino self-energy whiles the collision terms are 
obtained in the Born approximation. The resulting equations take 
the familiar form of the Boltzmann equation with corrections due to
the mixing both in the advection part and in the collision part. 
These corrections are essentially the same
as those derived by Sirera et al. for the advection terms and those by
Raffelt et al. for the collision terms, respectively,
though the formalism employed here is different from theirs. 
The derived equations will be
easily implemented in numerical codes employed in the simulations of 
supernova explosions and proto neutron star cooling.
\end{abstract}
%
\pacs{14.60.Pq 11.10.Wx 97.60.Bw 97.60.Jd}

\section{introduction}

The neutrino transport plays an important role in some astrophysical
phenomena such as supernova explosions and the following proto neutron 
star cooling (e.g. \cite{bc89,gr96,br85,su94} and references
therein). In their studies, 
the Boltzmann equation or its approximate versions are commonly
employed to describe the temporal variations of neutrino distributions in the
phase space. These equations are usually derived from 
the following assumptions\cite{lq66,mm89,sy99}: 
(1) the neutrinos are propagating along the geodesics for a massless 
particle, $p^{\mu}p_{\mu} = 0$, and the volume in the phase space occupied 
by these neutrinos is not varied along their world line if there is
no reaction. (2) the variation of the neutrino population due to
reactions is described by the so-called collision terms obtained with the
stohszahl ansatz. With the masses non-diagonal in the neutrino flavor
space, the neutrino oscillation occurs among different flavors of
neutrinos(e.g. \cite{fy94} and references therein). It is thus 
interesting from an academic 
point of view how this oscillation phenomenon is described by the 
generalized Boltzmann equations\cite{ru90,rs93a,rs93b,pa95,dn96,sp99}. It 
is also important from a practical point of view for those who are
interested in the possible significant consequences the oscillation
might give in astrophysical events\cite{lo95,be97,ks96,jr99}. In collapse-driven 
supernova explosions, for example, this is particularly the case if the
resonance of oscillation occurs near a  
neutrino sphere where neutrinos are interacting with other particles
and thus the oscillation should be treated simultaneously with these
reactions and possibly with the evolution of the matter distribution
as well. The purpose of this paper is to provide the formulation which
can be easily implemented in those numerical simulations.  
\par 
In considering the transport equation with the oscillation, we 
have to rely on a more formal derivation of the Boltzmann
equation. This might be done in a couple of ways. Sirera and P\'{e}rez\cite{sp99},
for instance, based their derivation on the relativistic Wigner
function approach in the mean field approximation. Although they 
took the relativistic kinematics properly into account, 
they did not obtain the collision terms, since it is difficult to 
go beyond the mean field approximation in
their formalism. Raffelt et al.\cite{rs93a,rs93b}, on the other hand, obtained their
transport equation via the density matrix approach. Although they
derived the collision terms, they did not consider the spatially
inhomogeneous system. In this paper, we derive the relativistic 
Boltzmann equation including corrections due to the oscillation both in 
the advection terms and the collision terms by employing the real time
formalism of the nonequilibrium field theory\cite{sw61,kd65,ch85}. 
In this approach, the dispersion relation and the collision 
terms are derived on the same basis, that is, a particular 
approximation for the self-energy of neutrinos, which is conveniently 
represented with Feynman diagrams. 
\par 
The paper is organized as follows. We first derive a generic form of
the transport equation without specifying particular equations of motion of 
fields. Then, the formulation is applied to the neutrino flavor
oscillations. In so doing, we ignore small corrections of the order 
of $m^{2}_{\nu}/E_{\nu}$ except for the terms responsible for the
flavor conversion, as is usually the case. Here 
$m_{\nu}$ and $E_{\nu}$ are typical mass and energy
of neutrinos in the observer's inertial frame. In this limit,
as shown later, the left handed neutrinos are decoupled from the right 
handed ones and the difference between Majorana mass and Dirac mass 
never shows up in the flavor mixing. The general relativistic
corrections are obtained up to the leading order of $\lambda_{\nu} / R$, 
where $\lambda_{\nu}$ is a typical wave length of neutrino and $R$ is
a scale height of the background matter distribution.

\section{formulation}

\subsection{general derivation of transport equations }

In this section we derive general transport equations for
multi-component fields based on the real time formalism of
nonequilibrium field theory by Keldysh\cite{sw61,kd65,ch85}. In this formalism, 
we introduce path-ordered products of operators on the closed
time-path, which extends from $t = - \infty $ to $t = + \infty $ then
back to $t = - \infty $. In this product, the operator with a time
argument which comes later on the above time-path is put to the left of 
other operators whose time arguments come earlier. Accordingly 
the path-ordered Green function is defined as
\begin{equation}
i \, G_{pij}(t_{1}, t_{2}) \ \equiv \ \langle \, T_{p} \ \psi _{i}(t_{1}) 
\ \psi _{j}^{\dagger}(t_{2}) \, \rangle \quad .
\end{equation} 
Here  $T_{p}$ stands for the path-ordered product of the
following operators. The subscript $i j$ of the Green function denotes 
the components of the field. The bracket $\langle \cdots \rangle $
represents that arguments are averaged over the ensemble specified by
a density operator $\rho$ as $T\!r\{\cdots \rho \}$, where $T\!r$ is a trace
operator. We define a generating functional of the Green function as
\begin{equation}
Z(J, J^{\dagger}) \ \equiv \ T\!r \left\{T_{p} \, \left[
\exp \left (i \, \sum _{i}
\int_{p} \! d^{4} \! x \ \left ( \, J^{\dagger }_{i}(x) \psi _{i}(x) 
\ + \  J_{i}(x) \psi ^{\dagger }_{i}(x) \right ) \right ) \right ]
\, \rho \right \} \ \equiv \ \exp \left [\, i \, W(J, J^{\dagger})\, \right ]
\quad .
\end{equation}
The Green function is obtained by the functional derivative,
$i \, G_{pij}(x, y) \, = \, 
\frac{\delta \ \ \ }{ i \, \delta \! J^{\dagger}_{i} (x)}
\frac{\delta \ \ \ }{ i \, \delta \! J^{\rule{0mm}{4pt}}_{j}(y)} 
\, Z(J, J^{\dagger}) |_{J,J^{\dagger} = 0}$.  
The generating functional for the connected Green function is denoted
as $W(J, J^{\dagger})$. Going to the interaction representation, we
obtain
\begin{eqnarray}
Z(J, J^{\dagger}) & = & T\!r \left\{T_{p} \, \left[
\exp \left (i \, \sum _{i}
\int_{p} \! d^{4} \! x \ \left ( \, J^{\dagger }_{i}(x) \psi 
_{\text{\tiny \it I}i}(x) 
\ + \  J_{i}(x) \psi _{\text{\tiny \it I}i}^{\dagger }(x) 
\ + \  {\cal L}_{int}(\psi _{\text{\tiny \it I}}(x), 
\psi _{\text{\tiny \it I}}^{\dagger }(x)) \right ) \right ) \right ]
\, \rho _{\text{\tiny \it I}} \right \} \nonumber \\
\label{eq:zint}
& = & \exp \left [\, i \, \int _{p} \! d^{4} \! y \ 
{\cal L}_{int} \left (\frac{\delta }{i\delta J^{\dagger}}, 
\frac{\delta }{i\delta J} \right ) \right ] \  
T\!r \left \{ \, T_{p} \, \left [ \,
\exp  \left (i \, \sum _{i}
\int_{p} \! d^{4} \! x \ \left ( \, J^{\dagger }_{i}(x) \psi 
_{\text{\tiny \it I}i}(x) 
\ + \  J_{i}(x) \psi _{\text{\tiny \it I}i}^{\dagger }(x) \right ) \right ) \right ]
\, \rho _{\text{\tiny \it I}} \right \} \quad ,
\end{eqnarray}
where the Lagrangian density for interactions is denoted as ${\cal L}_{int}$ and 
the subscript $I$ indicates that the variables are given in the
interaction representation. The last factor of the right hand side of 
Eq.~(\ref{eq:zint}) is the generating functional for the no interaction 
case, $Z_{0}(J, J^{\dagger})$, and is given as
\begin{equation}
Z_{0}(J, J^{\dagger})
\  = \ 
\label{eq:zno}
Z_{vac}(J, J^{\dagger}) \  T\!r \left \{ \, \text{\Large :} \,
\exp  \left (i \, \sum _{i}
\int_{p} \! d^{4} \! x \ \left ( \, J^{\dagger }_{i}(x) \psi 
_{\text{\tiny \it I}i}(x) 
\ + \  J_{i}(x) \psi _{\text{\tiny \it I}i}^{\dagger }(x) \right ) \right ) \,
\text{\Large :} \ \rho _{\text{\tiny \it I}} \right \}  \quad ,
\end{equation}
with the generating functional for vacuum, 
\begin{equation}
Z_{vac}(J, J^{\dagger}) \ = \ \exp \left ( - \, i \, \sum _{ij}
\int \!\! \int _{p} \! d^{4} \! x d^{4} \! y \ 
J_{j}(y) \, G^{0}_{pij}(x, y) \, J^{\dagger}_{i}(x) \right ) \quad . 
\end{equation}
Here $G^{0}_{pij}(x, y)$ is the path-ordered Green function for
vacuum. The normal order product is represented by 
$\text{\Large :} \cdots \text{\Large :}$ in Eq.~(\ref{eq:zno}). All the
information of the ensemble is included in the last term of
Eq.~(\ref{eq:zno}), $N_{p}(J, J^{\dagger}) = \exp [\,i \, 
W_{p}^{N}(J, J^{\dagger})\,]$. Its connected part 
$W_{p}^{N}(J, J^{\dagger})$ is in general expanded to cumulants as 
\begin{eqnarray}
W_{p}^{N}(J, J^{\dagger}) & = & 
\sum _{m, n = 1}^{\infty}\frac{1}{m! \, n!} \int \cdots \int _{p}
\! d^{4} \! y_{1} \cdots  d^{4} \! y_{n} \, 
d^{4} \! x_{1} \cdots  d^{4} \! x_{m} \ 
J(y_{1}) \cdots J(y_{n}) \, J^{\dagger}(x_{1}) \cdots J^{\dagger}(x_{m}) 
\nonumber \\ &&
\qquad\qquad\qquad\qquad\qquad\qquad\qquad\qquad\qquad\qquad 
\times \ W_{p}^{Nmn}(\, x_{1}, \cdots, x_{m} \ | \ y_{1}, \cdots, y_{n}) \quad .
\end{eqnarray}
In the following we assume that the expansion is terminated at the
quadratic order. This is true, for example, for the thermal
equilibrium  and the more general condition for this to be true can be 
found in the paper by Danielewicz\nolinebreak\cite{dz84}. 
With this assumption, we can expand
as usual the Green functions by the propagator which have corrections
originating from a particular ensemble.

The Dyson equations are obtained by the Legendre transformations:
\begin{equation}
\Gamma (\psi_{c}, \psi^{\dagger}_{c}) \ = \ 
W(J, J^{\dagger}) \ - \ J^{\dagger} \cdot \psi_{c} \ - \ 
J \cdot \psi_{c}^{\dagger} 
\end{equation}
with $\psi_{c}(x) = \frac{\delta\ \ \ \ }{\delta J^{\dagger}(x)}W$ and 
$\psi_{c}^{\dagger}(x) = \frac{\delta\ \ \ }{\delta J(x)}W$. We use the
abbreviation $J^{\dagger} \cdot \psi_{c} = \sum_{i}
\int _{p} \! d^{4} \! x J_{i}^{\dagger}(x)\psi_{ci}(x)$. Then the
following relations hold: 
$\delta \Gamma / \delta \psi_{c}(x) = -J^{\dagger}(x)$,
$\delta \Gamma / \delta \psi^{\dagger}_{c}(x) = -J(x)$. The  
Dyson equations take the integral form on the closed time-path as
\begin{equation}
\int _{p} \! d^{4} \! z \ G_{p}^{c}(x, z) \, \Gamma_{p}(z, y) \ = \ 
\int _{p} \! d^{4} \! z \ \Gamma_{p}(x, z) \, G_{p}^{c}(z, y) \ = \ 
\delta _{p}(x - y) \quad ,
\end{equation}
where the connected Green function and the vertex function are defined as 
$i \, G^{c}_{p}(x, y) \, = \, 
\frac{\delta \ \ \ }{ i \, \delta \! J^{\dagger} (x)}
\frac{\delta \ \ \ }{ i \, \delta \! J^{\rule{0mm}{4pt}}(y)} 
\, W(J, J^{\dagger})$ and 
$\Gamma_{p}(x, y) \, = \,
\delta ^{2} \Gamma \, / \, \delta \psi_{c}(y) \delta \psi_{c}^{\dagger}(x)
$, respectively. The $\delta$-function is extended on the closed
time-path as follows: $\delta _{p}(x - y) = \delta (x - y)$ for
$t_{x}$ and $t_{y}$ on the positive branch of the time-path extending from 
$t = - \infty $ to $t = + \infty $ and 
$\delta _{p}(x - y) = - \delta (x - y)$ for $t_{x}$ and $t_{y}$ 
on the negative branch of the time-path that runs from $t = + \infty $ to
$t = - \infty $. Introducing the matrix representations for the Green
function and the vertex function as
\begin{equation}
\hat{G} \ = \ \left (
\begin{array}[c]{cc}
G_{F} & G_{+}\\ G_{-} & G_{\bar{F}}
\end{array}
\right ) \quad , \qquad
\hat{\Gamma} \ = \ \left (
\begin{array}[c]{cc}
\Gamma_{F} & \Gamma_{+}\\ \Gamma_{-} & \Gamma_{\bar{F}}
\end{array}
\right ) \quad ,
\end{equation}
we can recast the Dyson equation in a single time representation:
\begin{equation}
\int \! d^{4} \! z \ \hat{G}^{c}(x, z) \, \sigma_{3} \, \hat{\Gamma}(z, y) \ = \ 
\int \! d^{4} \! z \ \hat{\Gamma}(x, z) \, \sigma_{3} \, \hat{G}^{c}(z, y) \ = \ 
\sigma_{3} \, \delta (x - y) \quad .
\end{equation}
In the above equations, the time integration runs from $t = - \infty$
to $t = + \infty$, and $\sigma_{3} = \left ( 
\begin{array}[c]{cc}
1 & 0\\ 0 & -1
\end{array}
\right )$ is the Pauli matrix. The subscripts $F$ and $\bar{F}$
indicate that the time 
arguments are both on the positive branch and on the negative branch,
respectively, while the subscript $+$ means that the first argument is 
located on the positive branch and the second on the negative branch,
and the subscript $-$ represents the other way around. It is clear
that $G_{F}$ is an ordinary Green function defined from the
chronologically ordered product while $G_{\bar{F}}$ is obtained from
the anti-chronological ordering. From these quantities, we further
define the retarded, advanced and correlation functions as
\begin{eqnarray}
G_{r} & \ = \ & G_{F} \ - \ G_{+} \quad ,\\
G_{a} & \ = \ & G_{F} \ - \ G_{-} \quad ,\\
G_{c} & \ = \ & G_{F} \ + \ G_{\bar{F}} \quad .
\end{eqnarray}
The counter parts for the vertex functions are defined in an analogous 
way. Using the identity 
$\Gamma_{F} + \Gamma_{\bar{F}} = \Gamma_{+} + \Gamma_{-}$, we can
express $\Gamma$'s in general as
\begin{eqnarray}
\Gamma_{\pm} & \ = \ & i\, (B \ \pm \ \ \, A) \ \ ,\\
\Gamma_{F} & \ = \ & \ \ \, D \ + \ i \, B \quad ,\\
\Gamma_{\bar{F}} & \ = \ & - \, D \ + \ i \, B \quad ,\\
\Gamma_{r} & \ = \ & \ \ \, D \ + \ i \, A \quad ,\\
\Gamma_{a} & \ = \ & \ \ \, D \ - \ i \, A \quad ,
\end{eqnarray}
where $A$, $B$ and $D$ are three Hermitian matrices. Solving the Dyson 
equations using these quantities, we obtain the general form of the
Green functions as 
\begin{eqnarray}
G_{r} & \ = \ & \quad \ \ \left [\, D \ + \ i \, A \, \right ]^{-1} \quad ,\\
G_{a} & \ = \ & \quad \ \ \left [\, D \ - \ i \, A \, \right ]^{-1} \quad ,\\
G_{\pm} & \ = \ & - \, i \, \left [\, D \ + \ i \, A \, \right ]^{-1} 
\left[ \, B \ \pm \ \ \, A \, \right ]\left [\, D \ - \ i \, A \, \right ]^{-1}
\quad ,\\
G_{F} & \ = \ & \quad \ \ \left [\, D \ + \ i \, A \, \right ]^{-1} 
\left[ \, D \ - \ i \, B \, \right ]\left [\, D \ - \ i \, A \, \right ]^{-1}
\quad ,\\
G_{\bar{F}} & \ = \ & - \ \ \left [\, D \ + \ i \, A \, \right ]^{-1} 
\left[ \, D \ + \ i \, B \, \right ]\left [\, D \ - \ i \, A \, \right ]^{-1}
\quad .
\end{eqnarray}
The dispersion relation is obtained from $D$ and $A$, while the 
distribution function is found from $B$ as shown shortly.

$D$, $A$ and $B$ can be represented in turn by the self-energy 
$\Sigma_{p}$ which is defined from the two point vertex function as
\begin{equation}
\Gamma_{p} \ = \ \Gamma_{p0} \ - \ \Sigma_{p} \quad ,
\end{equation}
where the free vertex function is $\Gamma_{p0}(x - y) =
S(\partial_{x}) \, \delta_{p}(x - y)$. Here the derivative operator is
taken from the free Lagrangian, ${\cal L}_{0} = \psi
^{\dagger} S(\partial) \, \psi$. Defining again the matrix components of
the self-energy in the single time representation, we obtain 
\begin{eqnarray}
\label{eq:defd}
D & \ = \ & S(\partial _{x}) \, \delta (x - y) \ - \ 
\frac{1}{2} \left( \, \Sigma_{F} \, - \, \Sigma_{\bar{F}} \, \right )\\
\label{eq:eqA}
A & \ = \ & \frac{1}{2} \, i \, \left ( \, \Sigma_{-} \, - \, 
\Sigma_{+} \, \right )\\
B & \ = \ & \frac{1}{2} \, i \, \left ( \, \Sigma_{-} \, + \, 
\Sigma_{+} \, \right ) \quad .
\end{eqnarray}
The self-energy, on the other hand, is given by the relation
\begin{equation}
\label{eq:slf}
\Sigma_{p}(x, y) \ = \ \left (
- \, i \, \left \langle \  T_{p} \  j(x) \ j^{\dagger}(y) 
\  \right \rangle \ - \ \delta _{p}(x \, - \, y) \, 
\left \langle \, \frac{\delta ^{2}\ \ \ \ \ \ }
{\delta \psi (y) \, \delta \psi ^{\dagger}(x)} \  
{\cal L}_{int} \, \right \rangle
\right )_{1PI} \quad ,
\end{equation} 
where the currents are defined as $j(x) = \frac{\delta \ \ \ }{\delta
\psi^{\dagger}(x)} \, {\cal L}_{int}$ and 
$j^{\dagger}(x) = \delta{\cal L}_{int} / \delta
\psi(x)$, and the subscript $1PI$ means the one particle irreducible
part. 

Now we introduce the distribution function, $n$. First we define
another Hermitian matrix $N$ from $B$ as
\begin{equation}
G_{c} \ = \  - \  \Gamma_{r}^{-1} \  2 \, i \, B \  \Gamma_{a}^{-1}
 \ = \  \Gamma_{r}^{-1} \, N \ - \  N \, \Gamma_{a}^{-1} .
\end{equation}  
Then it satisfies the following equation 
\begin{equation}
\label{eq:eqN}
N \, D \  - \ D \, N \ - \ i \, ( \, N \, A \ + \ A \, N \, ) \ = \ 
- \, 2 \, i \, B \quad .
\end{equation}
The matrix distribution function is finally defined as
\begin{equation}
N \ = \ 1 \ \mp \ 2 \, n \quad,
\end{equation}
where the upper and lower signs are taken for Fermion and Boson,
respectively. It is easily shown that this distribution function
becomes Fermi- or Bose-distribution function in the thermal
equilibrium case. In that case, $n$ can be simultaneously
diagonalized with $D$ and gives the distribution functions of  
quasi-particles. In general, however, $n$ has non-diagonal components 
even in the representation which diagonalizes $D$. These non-diagonal
components are responsible for the flavor mixing as discussed
below. Eq.~(\ref{eq:eqN}) gives the equation satisfied by $n$:
\begin{equation} 
n \, D \  - \ D \, n \ - \ i \, ( \, n \, A \ + \ A \, n \, ) \ = \ 
\pm \ i \, (\, B \, - \, A \, ) \ = \ \mp \ \Sigma_{+} \quad .
\end{equation}
Using Eq.~(\ref{eq:eqA}), we can rewrite the above equation,
\begin{equation}
\label{eq:eqn}
n \, D \  - \ D \, n \ = \ \frac{1}{2} \left [\,
(1 \, \mp \, n) \, (\, \mp \, \Sigma_{+} \, ) \ + \ 
(\, \mp \, \Sigma_{+} \, ) \, (1 \, \mp \, n) \, \right ]
\ - \ \frac{1}{2} \left [\, n \, \Sigma_{-} \ + \ \Sigma_{-} \, n
\, \right ] \quad .
\end{equation}
It is already clear that the right hand side of the above equation
describes collisional processes among the quasi-particles. 
In fact, $\left ( \pm \, i \, \Sigma_{+} \right )$ and 
$\left ( - \, i \, \Sigma_{-} \right )$ can be interpreted as the 
emission and absorption rates of quasi-particle. 

The transport equation as we know it is obtained by performing 
the so-called gradient expansion
for the above equation. The Wigner representation of a quantity 
$F(x, y)$ is obtained by making Fourier transformation with
respect to the relative coordinate as
\begin{equation}
F(k, X) \ = \ \int \!d^{4}(x - y) \ e^{ik(x - y)} \, F(x, y) \quad ,
\end{equation}
with the center of mass coordinate $X = (x + y) / 2$. The gradient
expansion is performed by taking the Wigner representation of both
sides of Eq.~(\ref{eq:eqn}) keeping only the leading
order of the derivative with respect to $X$. Thus, we obtain the transport
equation as
\begin{eqnarray}
\label{eq:gebo}
\frac{1}{2} \left [ \, \frac{\partial D(k, X)}{\partial k_{\mu}} \,
\cdot \, \frac{\partial n(k, X)}{\partial X^{\mu}} \right . & + &
\left . 
\frac{\partial n(k, X)}{\partial X^{\mu}} \, \cdot \, 
\frac{\partial D(k, X)}{\partial k_{\mu}} \, \right ] \  - \ 
\frac{1}{2} \left [ \, \frac{\partial D(k, X)}{\partial X^{\mu}} \,
\cdot \, \frac{\partial n(k, X)}{\partial k_{\mu}} \ + \ 
\frac{\partial n(k, X)}{\partial k_{\mu}} \, \cdot \, 
\frac{\partial D(k, X)}{\partial X^{\mu}} \, \right ] \nonumber \\
& - & i \, \left [ \, D(k, X) \, n(k, X) \ - \  n(k, X) \, D(k, X) 
\, \right ] \nonumber \\
& = & \frac{1}{2} \left \{ \, \left [1 \, \mp \, n(k, X) \, \right ]
\left [\, \mp \, i \, \Sigma_{+}(k, X) \, \right ] \ + \ 
\left [\, \mp \, i \, \Sigma_{+}(k, X) \, \right ]
\left [1 \, \mp \, n(k, X) \, \right ] \, \right \} \nonumber \\
& - & \frac{1}{2} \left \{ \, n(k, X) \, \left [ \, i \, 
\Sigma_{-}(k, X) \, \right ] \ + \ 
\left [ \, i \, \Sigma_{-}(k, X) \, \right ] \, n(k, X) \right \}
\quad .
\end{eqnarray}
It is evident that the first row of the above equation represents 
ordinary advection terms while the right hand side stands for the 
collision terms. The second row, on the other hand, does not
appear in the ordinary transport equation and we see below that this
term causes the mixing among neutrino flavors. What remains now to do is to
give the self-energy which determines not only the collision terms but 
also the dispersion relation, that is, $D$. We do this for the
neutrino mixing in the next section.

\subsection{neutrino transport equation with flavor mixings}  

We apply the general formulation obtained so far to the neutrino
transport. The following Lagrangian density is considered:
\begin{equation}
{\cal L} \ = \ \left \{
\begin{array}[c]{ll}
\frac{i}{2} \, \overline{\psi}_{\mbox{\tiny\it L}} \gamma^{\mu} \!\!
\stackrel{\leftrightarrow}{\partial}_{\mu} \!\! \psi_{\mbox{\tiny\it L}}
\ - \ \frac{1}{2} \, \overline{\psi}\rule[0cm]{0cm}{1ex}^{c}_{\mbox{\tiny\it L}} 
M_{\mbox{\tiny\it M}} \psi_{\mbox{\tiny\it L}}
\ - \ \frac{1}{2} \, \overline{\psi}_{\mbox{\tiny\it L}} 
M_{\mbox{\tiny\it M}}^{\dagger} \psi^{c}_{\mbox{\tiny\it L}}
\ + \ {\cal L}_{int}
& \mbox{for Majorana $\nu$}\\
\frac{i}{2} \, \overline{\psi}_{\mbox{\tiny\it L}} \gamma^{\mu} \!\!
\stackrel{\leftrightarrow}{\partial}_{\mu} \!\! \psi_{\mbox{\tiny\it L}}
\ + \ \frac{i}{2} \, \overline{\psi}_{\mbox{\tiny\it R}} \gamma^{\mu} \!\!
\stackrel{\leftrightarrow}{\partial}_{\mu} \!\! \psi_{\mbox{\tiny\it R}}
\ - \ \overline{\psi}_{\mbox{\tiny\it R}}M_{\mbox{\tiny\it D}} 
\psi_{\mbox{\tiny\it L}} \ - \ \overline{\psi}_{\mbox{\tiny\it L}}
M^{\dagger}_{\mbox{\tiny\it D}} \psi_{\mbox{\tiny\it R}}
\ + \ {\cal L}_{int} \qquad
& \mbox{for Dirac $\nu$}
\end{array} \right . \quad ,
\end{equation}
where the Majorana and Dirac masses are $M_{\mbox{\tiny\it M}}$ and 
$M_{\mbox{\tiny\it D}}$, respectively. The subscripts $L$ and $R$
stand for the spinor with left and right handed chirality,
respectively, and $\psi^{c}_{\mbox{\tiny\it L}} = C \overline{\psi}
\rule[0cm]{0cm}{1ex}^{\mbox{\tiny\it T}}_{\mbox{\tiny\it L}}$ with $C$
the charge conjugation and the superscript $T$ representing the
transposition. The interaction Lagrangian density is denoted as 
${\cal L}_{int}$. In the above equation, the indices for spinor
components and neutrino flavors are suppressed. In the following, the flavor is
denoted by the superscript and the spinor component by the subscript
as $\psi^{a}_{i}$ when necessary. 
The matrix Green functions of interest are 
$
\langle \, T_{p} \psi_{\mbox{\tiny\it L}}\rule[0cm]{0cm}{1ex}^{a}_{i}
\, \overline{\psi}_{\mbox{\tiny\it L}}\rule[0cm]{0cm}{1ex}^{b}_{j}\, \rangle
$, 
$
\langle \, T_{p} \, \psi_{\mbox{\tiny\it L}}\rule[0cm]{0cm}{1ex}^{a}_{i}
\, \overline{\psi}_{\mbox{\tiny\it R}}\rule[0cm]{0cm}{1ex}^{b}_{j}\, \rangle
$, 
$
\langle \, T_{p} \psi_{\mbox{\tiny\it R}}\rule[0cm]{0cm}{1ex}^{a}_{i}
\, \overline{\psi}_{\mbox{\tiny\it L}}\rule[0cm]{0cm}{1ex}^{b}_{j}\, \rangle
$ and 
$
\langle \, T_{p} \psi_{\mbox{\tiny\it R}}\rule[0cm]{0cm}{1ex}^{a}_{i}
\, \overline{\psi}_{\mbox{\tiny\it R}}\rule[0cm]{0cm}{1ex}^{b}_{j}\, \rangle 
$. Here and in the following, $\psi_{\mbox{\tiny\it R}}$ should be
replaced by $\psi_{\mbox{\tiny\it L}}^{c}$ for the Majorana neutrino. 
We discuss the advection part and collision part of the Boltzmann
equation separately, since we apply different approximations to the
self-energies included in them. 

\subsubsection{advection part}

\begin{figure}
\begin{center}
\epsfig{file=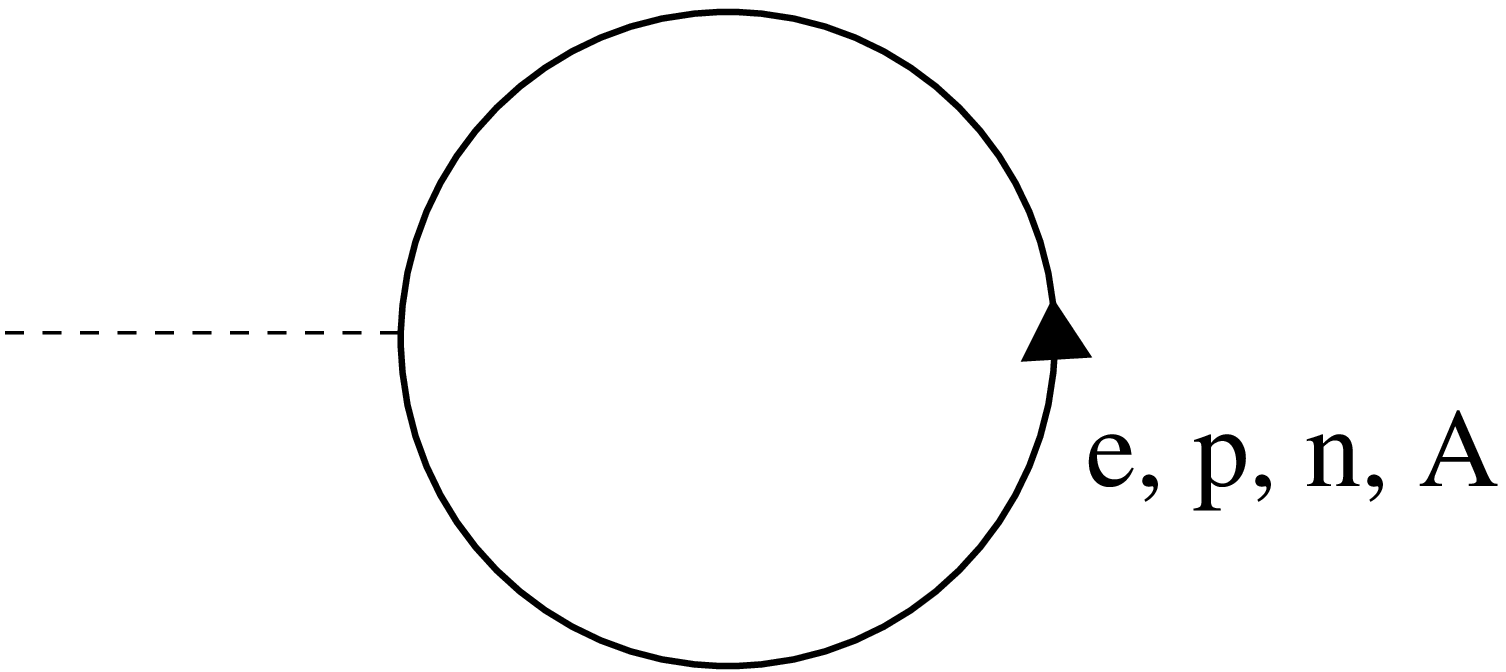, width=.50\textwidth}
\end{center}
\caption{The Feynman diagram for the self-energy of neutrino 
in the mean field approximation. The dashed line stands for the weak
interaction and the thick line represents the Green function for the
particles indicated in the figure.}
\label{fig1}
\end{figure}

Following the common practice, we take the mean field approximation 
for the neutrino self-energy in the advection part, which is 
conveniently represented by a Feynman
diagram shown in Fig.~\ref{fig1} and comes from the second term of 
Eq.~(\ref{eq:slf}). Only scattering processes contribute to this
ensemble average. In the supernova core, the scatterings on free
nucleons, nuclei and electrons are important. The former two of them
occur only via neutral currents and as a result, the self-energies
corresponding to them are proportional to the unit matrix in the
flavor space:
\begin{eqnarray}
{\cal L}_{int}^{nsc} & = & \sum_{a, N} - \frac{G_{F}}{\sqrt{2}}
\, \left [ \, \overline{\psi}_{\mbox{\tiny\it L}}\!\!\rule[0cm]{0cm}{1ex}^{a}
\, \gamma^{\mu}\left (1 \, - \, \gamma ^{5} \right ) 
\psi_{\mbox{\tiny\it L}}^{a} \, \right ]
\left [ \, \overline{\psi}_{\mbox{\tiny\it N}} \, 
\gamma_{\mu} \left (h_{\mbox{\tiny\it N}}^{\mbox{\tiny\it V}} \, - \, 
h_{\mbox{\tiny\it N}}^{\mbox{\tiny\it A}} \, \gamma ^{5} \right )
 \psi_{\mbox{\tiny\it N}} \, \right ] \quad ,\\
\Sigma ^{ab} _{F\mbox{\tiny\it LL}} & = & 
\hspace{1em} \delta ^{ab} \, \frac{G_{F}}{\sqrt{2}} \, \gamma^{\mu}
\left (1 \, - \, \gamma ^{5} \right ) \sum_{N}
h_{\mbox{\tiny\it N}}^{\mbox{\tiny\it V}} \, \rho_{\mbox{\tiny\it N}}(X)
\, \delta_{\mu 0} \quad ,\\
\label{eq:rreq}
\Sigma ^{ab} _{F\mbox{\tiny\it RR}} & = & 
- \, \delta ^{ab} \, \frac{G_{F}}{\sqrt{2}} \, \gamma^{\mu}
\left (1 \, + \, \gamma ^{5} \right ) \sum_{N}
h_{\mbox{\tiny\it N}}^{\mbox{\tiny\it V}} \, \rho_{\mbox{\tiny\it N}}(X)
\, \delta_{\mu 0} \quad ,\\
\Sigma ^{ab} _{F\mbox{\tiny\it LR}} & = & \Sigma ^{ab}_{F\mbox{\tiny\it RL}} 
\ = \ 0 \quad ,
\end{eqnarray} 
where Eq.~(\ref{eq:rreq}) is true only for the Majorana neutrino and 
$\Sigma ^{ab} _{F\mbox{\tiny\it RR}} = 0$ for the Dirac neutrino. In
the above equations, the subscript $N$ runs over  neutron and
proton, and $\rho_{\mbox{\tiny\it N}}$ stands for the nucleon number 
density. The similar equations are obtained for the scattering on
nuclei. Hence, in the following, the nucleon scattering is
considered. 

On the other hand, the scattering on electrons gives a 
non-trivial structure to the self-energy in the flavor space 
since the process occurs
not only through the neutral current but also through the charged
current, and the latter is relevant only for the electron-type
neutrinos in the matter in which electrons are abundant but other
charged leptons are not. In that case, the interaction Lagrangian
density becomes 
\begin{eqnarray}
{\cal L}_{int}^{esc} & = & - \, \frac{G_{F}}{\sqrt{2}} \, 
\left [ \, \overline{\psi}_{\mbox{\tiny\it L}}\!\!\rule[0cm]{0cm}{1ex}^{\nu _{e}}
\, \gamma ^{\mu} \left ( 1 \, - \, \gamma^{5} \right ) 
\psi_{\mbox{\tiny\it L}}^{\nu _{e}} \, \right ] 
\left [ \, \overline{\psi}_{e} \, \gamma_{\mu} \left
(\tilde{g}^{\mbox{\tiny\it V}} \, - \, \tilde{g}^{\mbox{\tiny\it A}} 
\, \gamma^{5} \right ) \psi_{e} \, \right ] \nonumber \\
& + & \sum_{a}^{\nu _{\mu}, \nu_{\tau}} - \, \frac{G_{F}}{\sqrt{2}} \, 
\left [ \, \overline{\psi}_{\mbox{\tiny\it L}}\!\!\rule[0cm]{0cm}{1ex}^{a}
\, \gamma ^{\mu} \left ( 1 \, - \, \gamma^{5} \right ) 
\psi_{\mbox{\tiny\it L}}^{a} \, \right ] 
\left [ \, \overline{\psi}_{e} \, \gamma_{\mu} \left
(g^{\mbox{\tiny\it V}} \, - \, g^{\mbox{\tiny\it A}} 
\, \gamma^{5} \right ) \psi_{e} \, \right ] \quad .
\end{eqnarray}
In the above equation, $g^{\mbox{\tiny\it V}} = 
- 1/2 + 2 \sin ^{2} \theta _{\mbox{\tiny\it W}}$
and $g^{\mbox{\tiny\it A}} - -1/2$ denote the vector and axial vector
coupling constants of the neutral current, while the charged current
is also taken into account in 
$\tilde{g}^{\mbox{\tiny\it V}} = g^{\mbox{\tiny\it V}} + 1$ and 
$\tilde{g}^{\mbox{\tiny\it A}} = g^{\mbox{\tiny\it A}} + 1$. The
Weinberg angle is referred to as $\theta _{\mbox{\tiny\it W}}$
here. We obtain the self-energy of neutrino in the mean field
approximation as 
\begin{eqnarray}
\Sigma^{ab}_{F\mbox{\tiny\it LL}} & 
\ = \ & \delta ^{a \nu_{e}} \delta^{b \nu_{e}} \, 
\frac{G_{F}}{\sqrt{2}} \, \gamma^{\mu} \left (1 \, - \, \gamma^{5}
\right ) \, \tilde{g}^{\mbox{\tiny\it V}} \, \rho_{e}(X) \, 
\delta _{\mu 0} \nonumber \\
& \ + \ & \delta ^{a \nu_{\mu , \tau}} \delta^{b \nu_{\mu , \tau}} \, 
\frac{G_{F}}{\sqrt{2}} \, \gamma^{\mu} \left (1 \, - \, \gamma^{5}
\right ) \, g^{\mbox{\tiny\it V}} \, \rho_{e}(X) \, 
\delta _{\mu 0} \quad ,
\end{eqnarray}
for the unpolarized electrons\cite{mh88,nr88}. 
Here the electron number density is denoted as 
$\rho_{e}$. As for the other components of the
self-energy, $\Sigma_{F\mbox{\tiny\it LR}} = \Sigma_{F\mbox{\tiny\it
RL}} = 0$ common to both types of neutrinos, and 
$\Sigma_{F\mbox{\tiny\it RR}} = - \Sigma_{F\mbox{\tiny\it LL}}$ with 
$(1 - \gamma^{5})$ replaced with $(1 + \gamma^{5})$ 
for the Majorana neutrino and 
$\Sigma_{F\mbox{\tiny\it RR}} = 0$ for the Dirac neutrino. If the electrons are
polarized in the magnetic field, the neutrino self-energy is modified
to\cite{ec96,eg96}
\begin{eqnarray}
\Sigma^{ab}_{F\mbox{\tiny\it LL}} 
& \ = \ & \delta ^{a \nu_{e}} \delta^{b \nu_{e}} \, 
\frac{G_{F}}{\sqrt{2}} \, \gamma^{\mu} \left (1 \, - \, \gamma^{5}
\right ) \left \{ \tilde{g}^{\mbox{\tiny\it V}} \, \rho_{e}(X) \, 
\delta _{\mu 0} \, - \, \tilde{g}^{\mbox{\tiny\it A}} \, \rho^{0}_{e}(X)
\, \delta _{\mu z} \right \} \nonumber \\
& \ + \ & \delta ^{a \nu_{\mu , \tau}} \delta^{b \nu_{\mu , \tau}} \, 
\frac{G_{F}}{\sqrt{2}} \, \gamma^{\mu} \left (1 \, - \, \gamma^{5}
\right ) \left \{ g^{\mbox{\tiny\it V}} \, \rho_{e}(X) \, 
\delta _{\mu 0} \, - \, g^{\mbox{\tiny\it A}} \, \rho^{0}_{e}(X)
\, \delta _{\mu z} \right \} \quad ,
\end{eqnarray}
where the magnetic field is parallel to the Z-axis. The electron
number density in the lowest Landau level is represented as $\rho^{0}_{e}$.
It is again true that the other components of the self-energy are zero 
except for $\Sigma_{F\mbox{\tiny\it RR}} = - \Sigma_{F\mbox{\tiny\it LL}}$ 
with $(1 - \gamma^{5}) \rightarrow  (1 + \gamma^{5})$ for the Majorana neutrino.
It is easily understood that neutrino-neutrino scatterings can be
treated just in the same way. 

Now that we obtain the specific form of the neutrino self-energy, we
can apply it to the left hand side of Eq.~(\ref{eq:gebo}). 
Suppressing the flavor and spinor indices and writing
only the chirality components in matrix form, we obtain $D$ in
Eq.~(\ref{eq:gebo}) using Eq.~(\ref{eq:defd}) as 
\begin{equation}
D \ = \ \left (
\begin{array}[c]{cc}
D_{\mbox{\tiny\it LL}} & D_{\mbox{\tiny\it LR}} \\
D_{\mbox{\tiny\it RL}} & D_{\mbox{\tiny\it RR}} 
\end{array} 
\right ) \ = \ \left (
\begin{array}[c]{cc}
k_{\mu} \gamma ^{\mu} \, - \, \Phi \gamma ^{0} \, - \, 
\Phi_{\mbox{\tiny\it B}} \gamma ^{z} & - M^{\dagger} \\
- M & k_{\mu} \gamma ^{\mu} \, + \, \Phi ' \gamma ^{0} \, + \, 
\Phi '_{\mbox{\tiny\it B}} \gamma ^{z}
\end{array}
\right ) \quad .
\end{equation}
Here the potentials are defined as
\begin{eqnarray}
\Phi & \ = \ & \ \ \, \delta ^{a \nu_{e}} \delta ^{b \nu_{e}} \, 
\sqrt{2} G_{F} \, \tilde{g}^{\mbox{\tiny\it V}} \, \rho_{e} \, + \, 
\delta ^{a \nu_{\mu , \tau }} \delta ^{b \nu_{\mu , \tau }} 
\sqrt{2} G_{F} \, g^{\mbox{\tiny\it V}} \, \rho_{e} \quad , \\
\Phi _{\mbox{\tiny\it B}} & \ = \ & - \, \delta ^{a \nu_{e}} \delta ^{b \nu_{e}}
\, \sqrt{2} G_{F} \, \tilde{g}^{\mbox{\tiny\it A}} \, \rho ^{0}_{e} \, - \,
\delta ^{a \nu_{\mu , \tau }} \delta ^{b \nu_{\mu , \tau }} 
\sqrt{2} G_{F} \, g^{\mbox{\tiny\it A}} \, \rho ^{0}_{e} \quad ,
\end{eqnarray} 
with $\Phi ' = \Phi$ and $\Phi '_{\mbox{\tiny\it B}} = 
\Phi _{\mbox{\tiny\it B}}$ for the Majorana neutrino, and $\Phi ' = 0$ 
and $\Phi '_{\mbox{\tiny\it B}} = 0$ for the Dirac neutrino. 
It is understood in the above equations that $\Phi _{\mbox{\tiny\it B}} 
= 0 $ in the case of no magnetic field. The dispersion relations for
quasi-particles are obtained from the eigen values of $D$. 

We first make an order estimate of each term in the
advection part. Defining $R$ as a typical length scale of the
matter distribution, the density scale height, for example, $E_{\nu}$ as a typical
energy of neutrino, and $\Delta m_{\nu}^{2}$ as a square mass difference, we 
find 
\begin{eqnarray}
\frac{\partial D}{\partial k} \cdot \frac{\partial n}{\partial X} 
& \ \sim \ & \frac{n}{R} \quad , \\
\frac{\partial D}{\partial X} \cdot \frac{\partial n}{\partial k}
& \ \sim \ & \frac{\Phi}{R} \cdot \frac{n}{E_{\nu}} \ \sim \ 
\frac{\Delta m_{\nu}^{2}}{E_{\nu}^{2}} \cdot \frac{n}{R} \quad , \\
D \cdot n & \ \sim \ & \frac{\Delta m_{\nu}^{2}}{E_{\nu}} \cdot n 
\ \sim \ \frac{R}{\lambda } \cdot \frac{n}{R} \quad .
\end{eqnarray}
Here $\lambda $ is a wave length corresponding to $\Delta m_{\nu}^{2}
/ E_{\nu}$ : $\lambda \sim 0.1 {\rm cm} 
[\Delta m_{\nu}^{2} / 1 {\rm eV}^{2}]^{-1} [E_{\nu} / 1 {\rm
MeV}]$. For the typical mass difference and energy of neutrino,
$\Delta m_{\nu}^{2} / E_{\nu}^{2} \sim 10^{-12}$. Hence the second 
term in the left hand side of Eq.~(\ref{eq:gebo}), which
represents the potential force exerted on neutrino by surrounding
matter, is much smaller than the first term, which corresponds
to the ordinary advection term in the Boltzmann equation. We ignore
the former in the following discussion.  

Next we show that the $n_{\mbox{\tiny\it LL}}$ can be decoupled from the  
other components assuming that $\Delta m_{\nu}^{2} / E_{\nu}^{2}$ is
neglected. We perform two matrix manipulations for 
$D \cdot n - n \cdot D$ : (1) multiply the first row with  
$k_{\mu} \gamma ^{\mu} \, + \, \Phi ' \gamma ^{0} \, + \, 
\Phi '_{\mbox{\tiny\it B}} \gamma ^{z}$ from the left and add to it the second 
row multiplied with $M^{\dagger}$ from the left. (2) then 
multiply the first column with 
$k_{\mu} \gamma ^{\mu} \, + \, \Phi ' \gamma ^{0} \, + \, 
\Phi '_{\mbox{\tiny\it B}} \gamma ^{z}$ from the right and add to it 
the second column multiplied with $M$ from the right. Taking into
account that $n_{\mbox{\tiny\it LR}} \sim n_{\mbox{\tiny\it RL}} \sim 
(\Delta m_{\nu} / E_{\nu}) \, n_{\mbox{\tiny\it LL}}$ and $\Phi / E_{\nu} 
\sim \Phi_{\mbox{\tiny\it B}} / E_{\nu} \sim \Delta m_{\nu}^{2} /
E_{\nu}^{2}$, we obtain the equation for $n_{\mbox{\tiny\it LL}}$ as
\begin{eqnarray}
\label{eq:eqmx}
\left [ \, k_{\mu}k^{\mu} \, - \, M^{2} \right . & - & 
\left . k_{\mu}\gamma^{\mu} \left (
\Phi \gamma^{0} \, + \, \Phi_{\mbox{\tiny\it B}} \gamma^{z} \right )
\, + \, \left ( \phi ' \gamma ^{0} \, + \, \Phi '_{\mbox{\tiny\it B}}
\gamma ^{z} \right ) k_{\mu} \gamma^{\mu} \, \right ] 
n_{\mbox{\tiny\it LL}} \, k_{\mu} \gamma^{\mu} \nonumber \\
& - & k_{\mu} \gamma^{\mu} \, n_{\mbox{\tiny\it LL}} 
\left [ \, k_{\mu}k^{\mu} \, - \, M^{2} \, + \, k_{\mu}\gamma^{\mu} \left (
\Phi ' \gamma^{0} \, + \, \Phi '_{\mbox{\tiny\it B}} \gamma^{z} \right )
\, - \, \left ( \Phi  \gamma ^{0} \, + \, \Phi _{\mbox{\tiny\it B}}
\gamma ^{z} \right ) k_{\mu} \gamma^{\mu} \, \right ] \quad .
\end{eqnarray}
The same manipulations are done for $\partial D / \partial k_{\mu}
\cdot \partial n / \partial X^{\mu} + \partial n / \partial X^{\mu}
\cdot \partial D / \partial k_{\mu}$ to obtain 
\begin{equation}
\label{eq:eqad}
k_{\mu} \gamma ^{\mu} \  \gamma^{\nu} \frac{\partial \,
n_{\mbox{\tiny\it LL}}}{\partial X ^{\nu}} \  k_{\sigma} \gamma^{\sigma} 
\ + \ k_{\sigma} \gamma^{\sigma} \  \frac{\partial \,
n_{\mbox{\tiny\it LL}}}{\partial X ^{\nu}} \gamma^{\nu} \  
k_{\mu} \gamma ^{\mu} \quad .
\end{equation}
In the next subsection, it is shown that $n_{\mbox{\tiny\it LL}}$ can 
be also separated from the other components in the collision terms by 
applying the same procedures.

To the leading order of $\Delta m_{\nu} / E_{\nu}$, 
$n_{\mbox{\tiny\it LL}}$ is a scalar with
respect to the spinor index. The familiar form of advection terms in the Boltzmann
equation is obtained by taking the trace with respect to the spinor
indices after multiplying Eqs.~(\ref{eq:eqmx}) and (\ref{eq:eqad}) with 
$\gamma^{0} (1 - \gamma ^{5}) / 2$ from the left :
\begin{equation}
\label{eqn:eqtr}
k^{\mu}\frac{\partial \, n^{ab}_{\mbox{\tiny\it LL}}}{\partial X^{\mu}}
\ + \ i k^{0} \left \{ \left ( \frac{M^{2\,ac}}{2k^{0}} \, + \, \Phi ^{ac}
\, + \, \frac{k^{z}}{k^{0}} \Phi^{ac}_{\mbox{\tiny\it B}} \right ) 
n^{cb}_{\mbox{\tiny\it LL}} \ - \ n^{ac}_{\mbox{\tiny\it LL}} 
\left (\frac{M^{2\,cb}}{2k^{0}} \, + \, \Phi ^{cb} \, + \, 
\frac{k^{z}}{k^{0}} \Phi^{cb}_{\mbox{\tiny\it B}} \right ) \right \}
\quad .
\end{equation}
Here the indices of flavor are explicitly included. From
this equation we see that the resultant equation is identical for the
Dirac and Majorana neutrinos up to the leading order of 
$\Delta m_{\nu} / E_{\nu}$ and that $D$ is effectively replaced by 
$D_{\mbox{\footnotesize eff}}$ in
the flavor space of the left handed neutrinos :
\begin{equation}
D_{\mbox{\footnotesize eff}} \ = \ \frac{1}{2} \left ( \, 
k_{\mu} k^{\mu} \ - \ M^{2} \ - \ 2 k^{0} \Phi 
\ - \ 2 k^{z} \Phi_{\mbox{\tiny\it B}}
\, \right ) \quad .
\end{equation}
The positive and negative zeros of $D_{\mbox{\footnotesize eff}}$
correspond to the energies of the neutrino and the anti-neutrino,
respectively. The transport equation for the anti-neutrino is obtained 
with the replacements : $k^{\mu} \rightarrow - \, k^{\mu}$, $n_{\mbox{\tiny\it LL}}
\rightarrow - \, n_{\mbox{\tiny\it LL}}$ in Eq.~(\ref{eqn:eqtr}). 
In the following, we
consider only the transport of on-shell neutrinos. Since we are not
interested in the small difference $\sim \Delta m_{\nu}^{2} / E_{\nu}$ 
of the on-shell energies among different flavors except in the terms 
responsible for the flavor mixing, we take $k^{0} = |\mbox{\boldmath
$k$}|$ in Eq.~(\ref{eqn:eqtr}).

In order to illuminate the structure of the advection part of the
transport equation obtained above, we discuss only the two-flavor
case of electron- and muon-neutrinos. Then Eq.~(\ref{eqn:eqtr})
multiplied with $i$ becomes on the flavor basis
\begin{equation}
\label{eqn:eqprz1}
i k^{\mu} \frac{\partial \, n^{ab}}{\partial X^{\mu}}
\ + \ k^{0} \, [ \, H , \ n \, ]^{ab} \quad , 
\end{equation}
with 
\begin{equation}
\label{eqn:eqprz2}
H \ = \ - \frac{\Delta M^{2}}{2 \, k^{0}} \ - \ \Phi \ - \ 
\frac{k^{z}}{k^{0}} \, 
\Phi_{\mbox{\tiny\it B}}
\ = \ \frac{1}{2 \, k^{0}} \left (
\begin{array}[c]{ll}
\displaystyle{\frac{\Delta _{0}}{2}} \cos 2 \theta _{0} & 
\ \ \, \displaystyle{\frac{\Delta _{0}}{2}} \sin 2 \theta _{0} \\
\displaystyle{\frac{\Delta _{0}}{2}} \sin 2 \theta _{0} &
- \, \displaystyle{\frac{\Delta _{0}}{2}} \cos 2 \theta _{0}
\rule[0cm]{0cm}{1.8em}
\end{array}
\right ) \ - \ \left (
\begin{array}[c]{cc}
\sqrt{2} G_{F} \rho_{e} \, - \, \displaystyle{\frac{k^{z}}{k^{0}}} \,
\sqrt{2} G_{F} \rho^{0}_{e} & 0 \\
0 & 0 \rule[0cm]{0cm}{1.8em}
\end{array}
\right ) \  ,
\end{equation}
where $\Delta M^{2}$ is a mass matrix with a diagonal matrix 
$(m_{1}^{2} + m_{2}^{2}) / 2 \cdot {\bf 1}$ subtracted. The eigen 
values of $M^{2}$ are $m_{1}^{2}$ and $m_{2}^{2}$ with the latter larger 
than the former, and their difference is defined to be 
$\Delta _{0} = m_{2}^{2} - m_{1}^{2}$. The mixing angle in vacuum is
denoted as $\theta _{0}$ and the following relation holds :
\begin{eqnarray}
M^{2} & \ = \ & U \, \left (
\begin{array}[c]{cc}
m_{1}^{2} & 0 \\ 0 & m_{2}^{2}
\end{array}
\right ) \, U^{\dagger} \quad ,\\
U & \ = \ & \left (
\begin{array}[c]{ll}
\cos \theta _{0} & - \, \sin \theta _{0} \\
\sin \theta _{0} & \ \ \ \cos \theta _{0}
\end{array}
\right ) \quad .
\end{eqnarray}
Eqs.~(\ref{eqn:eqprz1}) and (\ref{eqn:eqprz2}) are the same equations 
as obtained by Sirera and P\'{e}rez\cite{sp99}. 

Taking the bases on which $H$ in Eq.~(\ref{eqn:eqprz2}) is diagonalized at 
each point in space,
\begin{equation}
U^{\dagger}_{\mbox{\tiny \it M}}(x) \ H(x) \ U_{\mbox{\tiny \it M}}(x) \ = \ 
- \, \displaystyle{\frac{\tilde{M}^{2}(x)}{2 \, k^{0}}} \ \equiv \ 
\displaystyle{\frac{1}{2 \, k^{0}}} \left (
\begin{array}[c]{cc}
\tilde{M}^{2}_{1}(x) & 0  \\ 0 & \tilde{M}^{2}_{2}(x)
\end{array}
\right ) \quad ,
\end{equation}
with 
\begin{equation}
U_{\mbox{\tiny \it M}}(x) \ = \ \left (
\begin{array}[c]{ll}
\cos \theta _{\mbox{\tiny \it M}}(x) & 
- \, \sin \theta _{\mbox{\tiny \it M}}(x) \\
\sin \theta _{\mbox{\tiny \it M}}(x) & \ \ \ 
\cos \theta _{\mbox{\tiny \it M}}(x)
\end{array}
\right ) \quad ,
\end{equation}
we obtain
\begin{eqnarray}
U^{\dagger}_{\mbox{\tiny \it M}}(x) \ 
\left [ i k^{\mu} \frac{\partial \, n}{\partial X^{\mu}} \right ] \ 
U_{\mbox{\tiny \it M}}(x) & \ = & \ i k^{\mu} \frac{\partial \, \tilde{n}}
{\partial X^{\mu}} \ - \ \left [ \, \tilde{n} , \  
U^{\dagger}_{\mbox{\tiny \it M}}(x) \, 
i k^{\mu} \, \frac{\partial \, U_{\mbox{\tiny \it M}}(x)}
{\partial X^{\mu} \ \ \ \ } \, \right ]  \quad ,\\
& \ = & \ i k^{\mu} \frac{\partial \, \tilde{n}}
{\partial X^{\mu}} \ - \ \left [ \, \tilde{n} , \ 
\left ( 
\begin{array}[c]{rr}
0 & - 1 \\ 1 & 0 
\end{array}
\right )
i k^{\mu} \, \frac{\partial \, \theta_{\mbox{\tiny \it M}}(x)}
{\partial X^{\mu} \ \ \ \ } \, \right ] \quad .
\end{eqnarray}
Here the mass eigen values and mixing angle in matter are denoted as 
$\tilde{M}_{1}$, $\tilde{M}_{2}$ and $\theta_{\mbox{\tiny \it M}}$,
respectively, and the distribution function in this representation is 
defined as $\tilde{n}$. In order to see the oscillation among
different flavors, we write down each component of the above equation :
\begin{eqnarray}
\label{eqn:eqcom1}
i k^{\mu} \, \frac{\partial \, \tilde{n}^{11}}{\partial X^{\mu}}
& \ - \ & i  k^{\mu} \, \frac{\partial \, \theta_{\mbox{\tiny \it M}}(x)}
{\partial X^{\mu} \ \ \ \ } \left ( \, \tilde{n}^{12} \, + \,
 \tilde{n}^{21} \, \right ) \quad , \\
\label{eqn:eqcom2}
i k^{\mu} \, \frac{\partial \, \tilde{n}^{12}}{\partial X^{\mu}}
& \ + \ & i  k^{\mu} \, \frac{\partial \, \theta_{\mbox{\tiny \it M}}(x)}
{\partial X^{\mu} \ \ \ \ } \left ( \, \tilde{n}^{11} \, - \,
 \tilde{n}^{22} \, \right ) \ + \ \frac{\Delta 
_{\mbox{\tiny \it M}}}{2} \, \tilde{n}^{12} \quad ,\\
\label{eqn:eqcom3}
i k^{\mu} \, \frac{\partial \, \tilde{n}^{21}}{\partial X^{\mu}}
& \ - \ & i  k^{\mu} \, \frac{\partial \, \theta_{\mbox{\tiny \it M}}(x)}
{\partial X^{\mu} \ \ \ \ } \left ( \, \tilde{n}^{11} \, - \,
 \tilde{n}^{22} \, \right ) \ - \ \frac{\Delta 
_{\mbox{\tiny \it M}}}{2} \, \tilde{n}^{21} \quad , \\
\label{eqn:eqcom4}
i k^{\mu} \, \frac{\partial \, \tilde{n}^{22}}{\partial X^{\mu}}
& \ + \ & i  k^{\mu} \, \frac{\partial \, \theta_{\mbox{\tiny \it M}}(x)}
{\partial X^{\mu} \ \ \ \ } \left ( \, \tilde{n}^{12} \, + \,
 \tilde{n}^{21} \, \right ) \quad ,
\end{eqnarray}
where the mass square difference in matter is defined as $\Delta 
_{\mbox{\tiny \it M}} = \tilde{M}_{2}^{2} - \tilde{M}_{1}^{2}$. Ignoring the
collision terms for a moment and adding Eqs.~(\ref{eqn:eqcom1}) and 
(\ref{eqn:eqcom4}), we obtain the relation 
\begin{equation}
i k^{\mu} \, \frac{\partial \ \ }{\partial X^{\mu}} \, \left ( \, 
\tilde{n}^{11} \, + \, \tilde{n}^{22} \, \right ) \ = \ 0 \quad ,
\end{equation}
which expresses the number conservation of neutrinos. From
Eqs.~(\ref{eqn:eqcom2}) and (\ref{eqn:eqcom3}) the following equations 
are obtained,
\begin{eqnarray}
i k^{\mu} \, \frac{\partial \ \ }{\partial X^{\mu}} \, \left ( \, 
\frac{\tilde{n}^{12} \, + \, \tilde{n}^{21}}{2} \, \right ) & \, + \, & 
\frac{\Delta _{\mbox{\tiny \it M}}}{2} \, \left ( \,
\frac{\tilde{n}^{12} \, - \, \tilde{n}^{21}}{2} \, \right ) \ + \ 
i  k^{\mu} \, \frac{\partial \, \theta_{\mbox{\tiny \it M}}(x)}
{\partial X^{\mu} \ \ \ \ } \left ( \, \tilde{n}^{11} \, - \,
 \tilde{n}^{22} \, \right ) \  = \ 0 \quad , \\
i k^{\mu} \, \frac{\partial \ \ }{\partial X^{\mu}} \, \left ( \, 
\frac{\tilde{n}^{12} \, - \, \tilde{n}^{21}}{2} \, \right ) & \, + \, & 
\frac{\Delta _{\mbox{\tiny \it M}}}{2} \, \left ( \,
\frac{\tilde{n}^{12} \, + \, \tilde{n}^{21}}{2} \, \right )  \  = \  0 
\quad . 
\end{eqnarray}
Although we can infer the oscillating nature of the solution, it is better seen 
by eliminating $\tilde{n}^{12} \, - \, \tilde{n}^{21}$ and taking only the 
leading terms of $\Delta m_{\nu} / E_{\nu}$. 
The resultant equation roughly becomes 
\begin{equation}
\frac{d^{2}}{d\ell^{2}} \left ( \, 
\tilde{n}^{12} \, + \, \tilde{n}^{21} \, \right ) \ + \ 
\left ( \frac{\Delta _{\mbox{\tiny \it M}}}{2 E_{\nu}} \right )^{2}
\left ( \, \tilde{n}^{12} \, + \, \tilde{n}^{21} \, \right )
\ + \ \frac{1}{R^{2}} \, \left ( \, 
\tilde{n}^{11} \, - \, \tilde{n}^{22} \, \right )
\ \sim \ 0 \quad ,
\end{equation}
where $\ell$ is the path length and $R$ is the typical scale length of 
matter distribution. It is evident that 
$\tilde{n}^{12} \, + \, \tilde{n}^{21}$ have an oscillating part with
an oscillation length of $\lambda ^{-1} = 
\Delta _{\mbox{\tiny \it M}} \, / \, 2 \, E_{\nu}$ and a
non-oscillating part which is negligible when the adiabatic condition 
$\lambda / R  \ll 1$ is fulfilled. Hence we can ignore the
non-diagonal components $\tilde{n}^{12}$ and $\tilde{n}^{21}$ of 
the matrix distribution function if we are interested only in the variation of
the neutrino population on the length scale much longer than 
$\lambda $ and consider only the mass difference and energy of
neutrino which satisfy the above adiabatic condition, 
as is usually true for the supernova cores and proto
neutron stars. The non-diagonal components $\tilde{n}^{12}$ and 
$\tilde{n}^{21}$ can be ignored in the collision terms after taking
the average of the rapidly oscillating terms over the length scale much
larger than $\lambda $.

In the following, we set $\tilde{n}^{12} = \tilde{n}^{21} = 0$  and 
consider the equations governing the diagonal components of the
matrix distribution function for neutrinos. Following 
Raffelt et al.\cite{rs93b}, we 
represent $\tilde{n}^{11}$ and $\tilde{n}^{22}$ in terms of 
$n^{\nu_{e}}$ and $n^{\nu_{\mu}}$, the diagonal components on the 
flavor basis. From the relation 
\begin{equation}
U^{\dagger}_{\mbox{\tiny \it M}} \  n \  U_{\mbox{\tiny \it M}}
\ = \ \tilde{n} \ = \ \left (
\begin{array}[c]{cc}
\tilde{n}^{11} & 0 \\ 0 & \tilde{n}^{22}
\end{array}
\right ) \quad ,
\end{equation}
we obtain the distribution functions on the flavor basis as
\begin{equation}
\label{eq:eqadn}
n \ = \ \left (
\begin{array}[c]{cc}
n^{\nu_{e}} & \frac{1}{2} \tan 2\theta _{\mbox{\tiny \it M}}
\left ( \, n^{\nu_{e}} \, - \, n^{\nu_{\mu}} \, \right )\\
\frac{1}{2} \tan 2\theta _{\mbox{\tiny \it M}}
\left ( \, n^{\nu_{e}} \, - \, n^{\nu_{\mu}} \, \right )&
n^{\nu_{\mu}}
\end{array}
\right ) \quad .
\end{equation}
Inversely transforming the equation for $\tilde{n}$ in this approximation,
\begin{equation}
i k^{\mu} \, \frac{\partial \, \tilde{n}}{\partial X^{\mu}} 
\ = \ (\mbox{collision terms}) \quad ,
\end{equation}
we finally obtain the equation for $n$ in the limit of the adiabatic
oscillation between two flavors as
\begin{eqnarray}
\label{eq:eqfin}
i k^{\mu} \, \frac{\partial \, n}{\partial X^{\mu}} \ + \ 
\left \{ \tan 2 \theta _{\mbox{\tiny \it M}} \, 
\left ( \, n^{\nu_{e}} \, - \, n^{\nu_{\mu}} \, \right ) 
\left (
\begin{array}[c]{cc}
1 & 0 \\ 0 & -1
\end{array}
\right ) \right .
& \ - \ & \left . \left ( \, n^{\nu_{e}} \, - \, n^{\nu_{\mu}} \, \right ) 
\left (
\begin{array}[c]{cc}
0 & 1 \\ 1 & 0
\end{array}
\right ) \right \} \, i k^{\mu} \, 
\frac{\partial \, \theta _{\mbox{\tiny \it M}}(X)}
{\partial X^{\mu}\ \ \ \ } \nonumber \\
& \ = \ & (\mbox{collision terms}) \rule[0cm]{0cm}{1.5em} \quad .
\end{eqnarray}
Here the mixing angle in matter is given as $\tan 2 \theta
_{\mbox{\tiny \it M}} = \sin 2 \theta _{0} \, / \, \{ \,
\cos 2 \theta _{0} - 
[\, 2 \sqrt{2} \, G_{F} \, ( \, \rho_{e} -  k^{z} / k^{0} \, 
\rho^{0}_{e} \, ) \,
E_{\nu} \, ] / \Delta_{0} \, \}$.

We briefly discuss here the general relativistic correction
terms in the advection part. We take an arbitrary point in space-time 
and consider a small patch of space-time around it of the size of 
$\ell$ which satisfies the condition $R \gg \ell \gg 1/E_{\nu}$. 
Then we can take a local coordinates $X_{\mbox{\tiny \it R}}$ in this small region, 
which has a Minkowskian metric up to the second order of 
$\ell/R$. We also define an 
orthonormal tetrad $\mbox{\boldmath $e$}^{i}_{\mbox{\tiny \it R}}$ aligned to this 
coordinate and use it to project the four momentum of neutrino on it. 
On this coordinate in the small patch of space time, the above
derivation for the advection terms in the Boltzmann equation is still valid, 
that is, we obtain Eq.~(\ref{eqn:eqtr}) or Eq.~(\ref{eq:eqfin}) for the
adiabatic two-flavor mixing with trivial replacements of $X^{\mu}$ with 
$X^{\mu}_{\mbox{\tiny \it R}}$ and $k^{\mu}$ with 
$k^{\mu}_{\mbox{\tiny \it R}}$. Thus, there is no
additional mixing due to the general relativistic gravity under the
current assumption that the tiny mass difference of neutrinos is
ignored except for the mixing term, that is, the second terms of 
Eq.~(\ref{eqn:eqtr}) or Eq.~(\ref{eq:eqfin}). All we have to do now 
is to make a coordinate
transformation $X^{\mu}_{\mbox{\tiny \it R}} \rightarrow X^{\mu}$ and 
an associated momentum transformation $k^{\mu}_{\mbox{\tiny \it R}} 
\rightarrow k^{\mu}$. The latter is, 
in fact, induced by the transformation of the tetrads, 
$\mbox{\boldmath $e$}^{i}_{\mbox{\tiny \it R}} 
\rightarrow \mbox{\boldmath $e$}^{i}$,
the latter of which is given globally. Employing the orthogonal
transformation between two tetrads $\mbox{\boldmath $e$}^{i}(X) = 
T^{i}_{j}(X) \, \mbox{\boldmath $e$}^{j}_{\mbox{\tiny \it R}}$ 
and the transformation matrix
\begin{equation}
\left ( \begin{array}[c]{cc}
\displaystyle{\frac{\partial X^{\mu}}{\partial X^{\nu}_{\mbox{\tiny \it R}}}} & 
\displaystyle{\frac{\partial \, k^{j}}{\partial X^{\nu}_{\mbox{\tiny \it R}}}} \\
\displaystyle{\frac{\partial X^{\mu}}{\partial \, k^{i}_{\mbox{\tiny \it R}}}} &
\displaystyle{\frac{\partial \, k^{j}}{\partial \, k^{i}_{\mbox{\tiny \it R}}}}
\rule[0cm]{0cm}{4ex}
\end{array} \right ) \ = \ \left ( \begin{array}[c]{cc}
\displaystyle{\frac{\partial X^{\mu}}{\partial X^{\nu}_{\mbox{\tiny \it R}}}} & 
- \, 
\displaystyle{\frac{\partial X^{\rho}}{\partial X^{\nu}_{\mbox{\tiny \it R}}}} \, 
\displaystyle{\frac{\partial \, k^{m}_{\mbox{\tiny \it R}}}{\partial X^{\rho}}} \, 
\left ( \, T^{-1} \, \right )^{j}_{m} \\
0 & \left ( \, T^{-1} \, \right )^{j}_{i}
\rule[0cm]{0cm}{5ex}
\end{array} \right ) \quad ,
\end{equation}
we can perform the transformation for the advection term as follows:
\begin{eqnarray}
k^{i}_{\mbox{\tiny \it R}} \, \mbox{\boldmath $e$}_{\mbox{\tiny \it R}i}^{\mu} \, 
\frac{\partial \, n}{\partial X^{\mu}_{\mbox{\tiny \it R}}} & \ = \ & 
k^{i} \, \mbox{\boldmath $e$}_{i}^{\mu} \, 
\frac{\partial \, n}{\partial X^{\mu}} \ - \ k^{i} \, 
\mbox{\boldmath $e$}_{i}^{\mu} \, k^{m} \, 
\frac{\partial \, \langle \mbox{\boldmath $e$}_{m}
\cdot \, \mbox{\boldmath $e$}_{\mbox{\tiny \it R}}^{j} \rangle}
{\partial X^{\mu} \qquad \ \ \ }  \, \langle 
\mbox{\boldmath $e$}_{\mbox{\tiny \it R}j} \cdot \, 
\mbox{\boldmath $e$}_{n} \rangle 
\, \frac{\partial \, n}{\partial \, k^{n}} \quad , \nonumber \\
\label{eq:eqgr}
& = & k^{i} \, \mbox{\boldmath $e$}_{i}^{\mu} \, 
\frac{\partial \, n}{\partial X^{\mu}} \ - \ \omega ^{n}_{im} \, k^{i} 
\, k^{m} \, \frac{\partial \, n}{\partial \, k^{n}} \quad .
\end{eqnarray}
In the above equation, the inner product of two vectors is denoted as
$\langle \mbox{\boldmath $v$}_{1} \cdot \, \mbox{\boldmath $v$}_{2} 
\rangle$ and the component of the connection 1-form 
is designated as $\omega ^{k}_{ij} = \langle 
\nabla _{\!\!\mbox{\footnotesize \boldmath $e$}_{i}} \mbox{\boldmath $e$}_{j} \cdot 
\mbox{\boldmath $e$}^{k} \rangle$ with $\nabla _{\!\!\mbox{\footnotesize \boldmath
$e$}_{i}}$ the covariant derivative in the direction of $\mbox{\boldmath
$e$}_{i}$. The second term of Eq.~(\ref{eq:eqgr}) is a familiar
correction term due to the general relativity\cite{lq66,mm89,sy99}, which
accounts for the red shift and ray bending of neutrino in the
gravitational field. Since the mixing term and the collision terms (see 
below) do not contain spatial derivatives, they are unaffected by the
above transformation. However, $\partial \theta_{\mbox{\tiny \it M}} / 
\partial X^{\mu}_{\mbox{\tiny \it R}}$ in Eq.~(\ref{eq:eqfin}) is
affected just in the same way as $\partial n / \partial
X^{\mu}_{\mbox{\tiny \it R}}$ shown above. It is noted that this term
actually originates from the advection term $\partial n / \partial
X^{\mu}_{\mbox{\tiny \it R}}$ due to our pointwise choosing of 
the local mass eigen state basis. What remains to be done is, thus, 
to calculate the connection 1-form.

\subsubsection{collision part}

\begin{figure}
\begin{center}
\epsfig{file=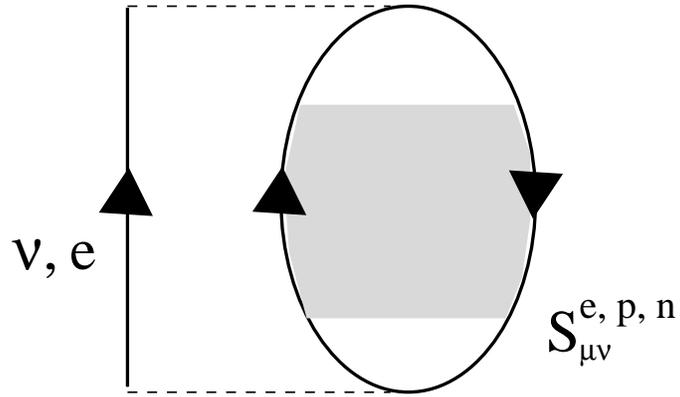,width=.50\textwidth}
\end{center}
\caption{The Feynman diagram for the self-energy of neutrino 
in the Born approximation.The dashed line stands for the weak
interaction and the thick lines represent the Green functions for the
particles indicated in the figure. The part with a shaded region 
denotes the structure functions (see the text for definitions) symbolically.}
\label{fig2}
\end{figure}

In this section, we derive collision terms in the Born approximation
for the neutrino self-energy. It is well known that the approximation
of the self-energy for the advection terms is different from that for
the collision terms\cite{kb62}. The Born approximation is conveniently
represented by the Feynman diagram shown in Fig.~\ref{fig2}. Only the
first term of the right hand side of Eq.~(\ref{eq:slf}) contributes to 
$\Sigma_{\pm}$ in the collision part. As done in the previous section, 
we evaluate the self-energy coming from various processes separately in
the following.

For the nucleon scattering, the self-energies are obtained as
\begin{eqnarray}
i \, \Sigma ^{ab} _{-\mbox{\tiny\it LL}}(x, y) & \ = \ & 
\, \frac{G_{F}^{2}}{2} \, \left [ \, \gamma^{\mu}
\left (1 \, - \, \gamma ^{5} \right ) i \, 
G^{ab}_{-\mbox{\tiny\it LL}}(x, y) \, \gamma^{\nu}
\left (1 \, - \, \gamma ^{5} \right ) \, \right ]
S_{\mbox{\tiny\it N}\mu \nu} (x, y) \quad ,\\
i \, \Sigma ^{ab} _{+\mbox{\tiny\it LL}}(x, y) & \ = \ & 
\, \frac{G_{F}^{2}}{2} \, \left [ \, \gamma^{\mu}
\left (1 \, - \, \gamma ^{5} \right ) i \, 
G^{ab}_{+\mbox{\tiny\it LL}}(x, y) \, \gamma^{\nu}
\left (1 \, - \, \gamma ^{5} \right ) \, \right ]
S_{\mbox{\tiny\it N}\nu \mu} (y, x) \quad .
\end{eqnarray}
In the above equations, the dynamical structure function for nucleon
is defined with the weak neutral current of nucleon 
$J_{\mbox{\tiny\it N}}^{\mu}(x)$ as 
$S_{\mbox{\tiny\it N}}^{\mu \nu} (x, y) = \langle 
J_{\mbox{\tiny\it N}}^{\mu}(x) J_{\mbox{\tiny\it N}}^{\nu}(y) \rangle$.
The weak neutral current for nucleon is given by $J_{\mbox{\tiny\it
N}}^{\mu} = \overline{\psi}_{\mbox{\tiny\it N}} \, 
\gamma_{\mu} \left (h_{\mbox{\tiny\it N}}^{\mbox{\tiny\it V}} \, - \, 
h_{\mbox{\tiny\it N}}^{\mbox{\tiny\it A}} \, \gamma ^{5} \right )
 \psi_{\mbox{\tiny\it N}} $. The other components of the matrix Green
function are zero for the Dirac neutrinos. For the Majorana neutrinos, 
$\Sigma_{\mbox{\tiny\it LR}}$ is obtained, for example, by replacing 
$G_{\mbox{\tiny\it LL}}$ with $G_{\mbox{\tiny\it LR}}$ and 
$\gamma^{\nu}\left (1 \, - \, \gamma ^{5} \right )$ with 
$\gamma^{\nu}\left (1 \, + \, \gamma ^{5} \right )$.
Recalling the relations $n_{\mbox{\tiny\it LR}} \sim n_{\mbox{\tiny\it RL}} \sim 
(\Delta m_{\nu} / E_{\nu}) \, n_{\mbox{\tiny\it LL}} \sim 
(\Delta m_{\nu} / E_{\nu}) \, n_{\mbox{\tiny\it RR}}$ and 
$\Sigma_{\mbox{\tiny\it LR}} \sim \Sigma_{\mbox{\tiny\it RL}} \sim 
(\Delta m_{\nu} / E_{\nu}) \Sigma_{\mbox{\tiny\it LL}} \sim 
(\Delta m_{\nu} / E_{\nu}) \Sigma_{\mbox{\tiny\it RR}}$, we find that
the $LL$-component can be decoupled from the other components after
the same matrix manipulations as done for the advection part and that
the resulting collision terms are identical for the Dirac and Majorana 
neutrinos, if we
take only the leading terms of $\Delta m_{\nu} / E_{\nu}$. Note that
the exchanged terms are added in the collision part unlike in the
advection part. From the term $n \cdot i \Sigma_{-}$, for example, we obtain 
$\left [ \, k _{\mu} \gamma ^{\mu} \, n_{\mbox{\tiny\it LL}} \, \right ]
\cdot \left [\, i \Sigma_{-\mbox{\tiny\it LL}} \, k _{\mu} \gamma ^{\mu} 
\, \right ]$ as the $LL$-component.

Following the procedures taken for the advection part, we 
multiply the collision terms with 
$\gamma ^{0} \left (1 - \gamma ^{5} \right ) / 2$ from the left, 
take the trace with respect to the spinor indices and divide by 
$4 k^{0}$. We then obtain from $n \cdot i \Sigma_{-}$, for example, the 
following:
\begin{equation}
\label{eq:eqcol1}
\int \! \! \frac{d^{4}q}{(2 \pi )^{4}} \frac{1}{4 k^{0}} \  Tr \left \{
\gamma^{0} \, \frac{1 - \gamma ^{5}}{2} \   k_{\rho} \gamma^{\rho} \ 
n_{\mbox{\tiny\it LL}} \, \frac{G_{F}^{2}}{2} \  \gamma^{\mu} \left (
1 - \gamma^{5} \right ) \ i G_{-\mbox{\tiny\it LL}}(k - q, X)\ 
\gamma^{\nu}\left (1 - \gamma^{5} \right ) \ k_{\sigma} \gamma^{\sigma} \right 
\} S_{\mbox{\tiny\it N}\mu \nu} (q, X) \quad .
\end{equation}
Ignoring again the tiny masses of neutrinos and $A$ in deriving $G_{\pm}$
from $D$, $A$ and $n$, we obtain 
\begin{equation}
\label{eq:eqgpm}
i G_{\pm \mbox{\tiny\it LL}}(k) \ \sim \ \frac{1 - \gamma ^{5}}{2} \ 
k_{\mu} \gamma^{\mu} \ 2 \pi \ \delta(k^{2}) \ 
\left [ \, \Theta (k^{0}) \left \{
\begin{array}[c]{c}
- \, n_{\nu} (\mbox{\boldmath $k$}) \\ 1 \, - \, n_{\nu}(\mbox{\boldmath $k$})
\end{array} \right \} \ + \ \Theta (-k^{0}) \left \{
\begin{array}[c]{c}
1 \, - \, n_{\bar{\nu}} (\mbox{\boldmath $k$})\\ - \, 
n_{\bar{\nu}} (\mbox{\boldmath $k$})
\end{array} \right \} \right ] \quad .
\end{equation} 
In the above equation, it is explicitly indicated that the negative
energy contribution to the number density of the neutrino corresponds to
the number density of the anti-neutrino. It is noted that the number
density is a function of $\mbox{\boldmath $k$}$ after we ignored $A$ and imposed an
on-shell condition $k^{2} = 0$. The upper (lower) components
in the columns correspond to $G_{+}$ ($G_{-}$). Inserting this relation to 
Eq.~(\ref{eq:eqcol1}) and recalling that $n$ is a scalar with respect
to the spinor indices, we obtain the following collision term for the
neutrino distribution function:
\begin{equation}
\label{eq:eqcol2}
\int \!\! \frac{d^{3}k'}{(2\pi)^{3}} \ \frac{1}{2 k^{'0}} \ \frac{1}{2}
\ n_{\nu}(k) \left [\, 1 \, - \, 
n_{\nu}(k') \, \right ] \ \frac{G_{F}^{2}}{2} 
\ L^{\mu \nu}(k, k') \, S_{\mbox{\tiny\it N}\mu \nu} (q, X) \quad ,
\end{equation}
where $k' = k - q$ is the four momentum of the scattered neutrino, and 
the tensor $L^{\mu \nu}$ is given as 
\begin{equation}
L^{\mu \nu}(k, k') \ = \ Tr \left \{
k_{\rho}\gamma^{\rho} \, \gamma^{\mu} (\, 1 \, - \, \gamma^{5} \, )
\, k'_{\sigma} \gamma^{\sigma} \, \gamma^{\nu} 
(\, 1 \, - \, \gamma^{5} \, ) \right \}
\ = \ 8 \, \left \{ k^{\mu}k^{'\nu} \, + \, k^{\nu}k^{'\mu} \, - \, 
g^{\mu \nu} k^{\rho}k'_{\rho} \, - \, i 
\varepsilon^{\mu\nu\rho\sigma}k_{\rho}k'_{\sigma} \right \} \  .
\end{equation}
Here the metric tensor is denoted as $g^{\mu\nu}$ and the
anti-symmetric tensor as $\varepsilon^{\mu\nu\rho\sigma}$ with 
$\varepsilon^{0123} = 1$. From the term $i \Sigma_{-} \cdot n$ we
obtain the collision term which is obtained from Eq.~(\ref{eq:eqcol2}) 
by replacing $n_{\nu}(k) \left [\, 1 \, - \, 
n_{\nu}(k') \, \right ]$ with $\left [\, 1 \, - \, 
n_{\nu}(k') \, \right ] n_{\nu}(k)$. 
Just in the same way we obtain from the term 
$[1 - n] [- i \Sigma_{+}]$ the following collision term:
\begin{equation}
\label{eq:eqcol3}
\int \!\! \frac{d^{3}k'}{(2\pi)^{3}} \ \frac{1}{2 k^{'0}} \ 
\frac{1}{2} \  \left [\, 1 \, - \, 
n_{\nu}(k) \, \right ] n_{\nu}(k') \ \frac{G_{F}^{2}}{2} 
\ L^{\mu \nu}(k, k') \, S_{\mbox{\tiny\it N}\nu \mu} (- q, X) \quad .
\end{equation}
For the term $[- i \Sigma_{+}][1 - n]$ we replace $\left [\, 1 \, - \, 
n_{\nu}(k) \, \right ] n_{\nu}(k')$ with $n_{\nu}(k') \left [\, 1 \, - \, 
n_{\nu}(k) \, \right ]$ in the above equation. Using the relation 
$S_{\mbox{\tiny\it N}\nu \mu} (- q) = e^{-\beta q^{0}} 
S_{\mbox{\tiny\it N}\mu \nu} (q)$ for the matter in equilibrium, 
which stands for the detailed balance, we finally obtain the collision 
terms as
\begin{eqnarray}
\int \!\! \frac{d^{3}k'}{(2\pi)^{3}} \ \frac{1}{2 k^{'0}} 
& \ \displaystyle{\frac{1}{2}} \ & \left \{ \, e^{-\beta q^{0}} \ 
\frac{n_{\nu}(k') \left [\, 1 \, - \, 
n_{\nu}(k) \, \right ] \, + \, 
\left [\, 1 \, - \, n_{\nu}(k) \, \right ] n_{\nu}(k')}{2} \right .
\nonumber \\ 
\label{eq:eqcol4}
& & \quad \ - \ \ \ \ \left . 
\frac{n_{\nu}(k) \left [\, 1 \, - \, 
n_{\nu}(k') \, \right ] \, + \, 
\left [\, 1 \, - \, n_{\nu}(k') \, \right ] n_{\nu}(k)}{2} \, \right \}
\ \frac{G_{F}^{2}}{2} 
\ L^{\mu \nu}(k, k') \, S_{\mbox{\tiny\it N}\mu \nu} (q, X) \quad .
\end{eqnarray}
If the matrix distribution function is diagonal, the above equation reduces to
the ordinary collision term.

If the mixing occurs adiabatically, the above collision term is 
further simplified. For the two-flavor case ($\nu_{e}$ and
$\nu_{\mu}$, for example), we insert the matrix 
distribution function given by Eq.~(\ref{eq:eqadn}) into 
Eq.~(\ref{eq:eqcol4}). Then we obtain the collision term for the 
$\nu_{e}$ distribution function as
\begin{eqnarray}
\int \!\! \frac{d^{3}k'}{(2\pi)^{3}} \ \frac{1}{2 k^{'0}} 
& \ \displaystyle{\frac{1}{2}} \ & \left \{ \, e^{-\beta q^{0}}   
\left [ \, n^{\nu_{e}}(k') \left [\, 1  -  
n^{\nu_{e}}(k) \, \right ] \, - \, \frac{1}{2} \, 
\tan 2\theta_{\mbox{\tiny \it M}}(k') \left [ \, 
n^{\nu_{e}}(k')  -  n^{\nu_{\mu}}(k') \, \right ] \,
\frac{1}{2} \, \tan 2\theta_{\mbox{\tiny \it M}}(k) \left [ \, 
n^{\nu_{e}}(k)  -  n^{\nu_{\mu}}(k) \, \right ] \,
\right ] \right .
\nonumber \\
& & \quad \ - \ \ \ \, \left . \left [ \,
n^{\nu_{e}}(k) \left [\, 1  -  
n^{\nu_{e}}(k') \, \right ] \, - \, 
\frac{1}{2} \, 
\tan 2\theta_{\mbox{\tiny \it M}}(k) \left [ \, 
n^{\nu_{e}}(k)  -  n^{\nu_{\mu}}(k) \, \right ] \,
\frac{1}{2} \, \tan 2\theta_{\mbox{\tiny \it M}}(k') \left [ \, 
n^{\nu_{e}}(k')  -  n^{\nu_{\mu}}(k') \, \right ] \,
\right ] \right \}
\nonumber \\
\label{eq:eqcol5}
& \times & \quad \frac{G_{F}^{2}}{2} 
\ L^{\mu \nu}(k, k') \, S_{\mbox{\tiny\it N}\mu \nu} (q, X) \quad .
\end{eqnarray}
The collision term for the $\nu_{\mu}$ distribution function is obtained 
by replacing $n^{\nu_{e}}$ with $n^{\nu_{\mu}}$ in the above equation. 
It is noted that the correction terms due to mixing cancel each other 
for iso-energetic scatterings, which we commonly assume for the 
neutrino-nucleon scattering in the supernova cores and proto neutron 
stars\cite{br85,sc90,yt99}.

The collision terms for the neutrino-electron scattering are essentially 
the same as those obtained for the nucleon scattering. The main difference 
originates from the fact that the electron weak current has flavor dependence, 
which gives rise to non-trivial contractions of flavor indices between the electron 
structure function and the neutrino distribution function such as 
$n_{\nu}^{ac}(k) \left [ 1 - n_{\nu}^{cb} \right ] S_{e}^{cb}$, where the 
superscripts $a, b, c$ represent flavors. The structure function $S_{e}$ is 
an electron counter part of the nucleon structure function 
$S_{\mbox{\tiny \it N}}$ and is defined as 
$S_{e\mu\nu}^{ab}(x, y) = \langle J_{e\mu}^{a}(x) J_{e\nu}^{b}(y) \rangle$. 
Here the weak current for electron is given as $J_{e}^{\mu} = 
\overline{\psi}_{e} \, \gamma^{\mu} \left (\tilde{g}^{\mbox{\tiny\it V}} \, - \, 
\tilde{g}^{\mbox{\tiny\it A}} \, \gamma ^{5} \right ) \psi_{e} $ for 
the $\nu_{e}$ scattering and $J_{e}^{\mu} = 
\overline{\psi}_{e} \, \gamma^{\mu} \left (g^{\mbox{\tiny\it V}} \, - \, 
g^{\mbox{\tiny\it A}} \, \gamma ^{5} \right ) \psi_{e} $ for the $\nu_{\mu}$ 
and $\nu_{\tau}$ scatterings, respectively. As a result, the collision 
term for the neutrino-electron scattering becomes for the electron
type neutrino in the case of the adiabatic two-flavor mixing as
\begin{eqnarray}
\int \!\! \frac{d^{3}k'}{(2\pi)^{3}} \ \frac{1}{2 k^{'0}} 
& \ \displaystyle{\frac{1}{2}} \ & \left \{ \rule[0cm]{0cm}{1.5em}
\left [ \, e^{-\beta q^{0}}   
\, n^{\nu_{e}}(k') \left [\, 1 \, - \,  
n^{\nu_{e}}(k) \, \right ] \, - \, n^{\nu_{e}}(k) \left [\, 1 \, - \, 
n^{\nu_{e}}(k') \, \right ] \right ] \right . 
S^{\nu_{e}\nu_{e}}_{e\mu\nu}(q, X)\nonumber \\
& \ - \ & \quad \left [ \, e^{-\beta q^{0}} \, - \, 1 \, \right ]
\frac{1}{2} \, \tan 2\theta_{\mbox{\tiny \it M}}(k') \left [ \, 
n^{\nu_{e}}(k')  \, - \, n^{\nu_{\mu}}(k') \, \right ] \,
\frac{1}{2} \, \tan 2\theta_{\mbox{\tiny \it M}}(k) 
\left [ \, n^{\nu_{e}}(k)  -  n^{\nu_{\mu}}(k) \, \right ] \nonumber \\
& \ \times \ & \quad \left . \frac{S^{\nu_{e}\nu_{\mu}}_{e\mu\nu}(q, X) 
\, + \, S^{\nu_{\mu}\nu_{e}}_{e\mu\nu}(q, X)}{2} \right \}
\nonumber \\
\label{eq:eqcol6}
& \times & \quad \frac{G_{F}^{2}}{2} 
\ L^{\mu \nu}(k, k') \quad .
\end{eqnarray}
The $\nu_{\mu}$ counterpart is obtained by the replacement of
$\nu_{e} \leftrightarrow \nu_{\mu}$ in the flavor indices in the above
equation.  

Although we assumed in the above derivation that the four momentum 
transfer is space-like to describe the scatterings, it is obvious that
the same Feynman diagram represents the annihilation and creation 
of neutrino pairs if  the transferred four momentum
is time-like. As stated above, the transport equations for the
anti-neutrinos are obtained from the negative energy part of the
distribution function. in that case, it is noted that the mixing angle
$\theta_{\mbox{\tiny \it M}}$ should be also calculated for the
negative energy. As is obvious from Eq.~(\ref{eqn:eqprz2}) the
sign of the potentials is changed for the anti-neutrinos and the
resonance conversion does not occur in this case as is well known. 
It is noted that neutrino-neutrino scatterings are treated just in the 
same way by substituting the neutrino structure function, which in
turn should be evaluated with the neutrino Green functions, 
Eq.~(\ref{eq:eqgpm}).

Next we consider the neutrino emission and absorption on
nucleons. For the temperature and neutrino energy of current interest, 
the muon is not abundant and only the electron-type neutrino is
involved in this process. The interaction Lagrangian density is 
\begin{equation}
{\cal L}^{abs}_{int} \ = \ - \frac{G_{F}}{\sqrt{2}}
\, \left [ \, \overline{\psi}_{\mbox{\tiny\it e}}
\, \gamma^{\mu}\left (1 \, - \, \gamma ^{5} \right ) 
\psi_{\mbox{\tiny\it L}}^{\nu_{e}} \, \right ]
\left [ \, \overline{\psi}_{\mbox{\tiny\it p}} \, 
\gamma_{\mu} \left (g^{\mbox{\tiny\it V}} \, - \, 
g^{\mbox{\tiny\it A}} \, \gamma ^{5} \right )
 \psi_{\mbox{\tiny\it n}} \, \right ] \ + \ H.\,C. \quad ,
\end{equation}
where the coupling constants $g^{\mbox{\tiny\it V}} = 1.0$ and 
$g^{\mbox{\tiny\it A}} = 1.23$, and $H.\,C.$ stands for the Hermite
conjugate. In the Born approximation, the self-energy is given by
\begin{eqnarray}
i \, \Sigma ^{\nu_{e}\nu_{e}} _{-\mbox{\tiny\it LL}}(x, y) & \ = \ & 
\, \frac{G_{F}^{2}}{2} \, \left [ \, \gamma^{\mu}
\left (1 \, - \, \gamma ^{5} \right ) i \, 
G^{e}_{-}(x, y) \, \gamma^{\nu}
\left (1 \, - \, \gamma ^{5} \right ) \, \right ]
S_{\mbox{\tiny\it pn}\mu \nu} (x, y) \quad ,\\
i \, \Sigma ^{\nu_{e}\nu_{e}} _{+\mbox{\tiny\it LL}}(x, y) & \ = \ & 
\, \frac{G_{F}^{2}}{2} \, \left [ \, \gamma^{\mu}
\left (1 \, - \, \gamma ^{5} \right ) i \, 
G^{e}_{+}(x, y) \, \gamma^{\nu}
\left (1 \, - \, \gamma ^{5} \right ) \, \right ]
S_{\mbox{\tiny\it np}\nu \mu} (y, x) \quad ,
\end{eqnarray}
where the Green functions for electrons are denoted as $G^{e}_{\pm}$
and the structure functions for the charged weak currents of nucleons 
are defined, for example, as $S^{\mu\nu}_{pn}(x, y) = 
\langle J^{\mu}_{pn}(x) \, J^{\nu}_{np}(y) \rangle $ with the charged
current given by $J^{\mu}_{np} = \overline{\psi}_{\mbox{\tiny\it p}} \, 
\gamma_{\mu} \left (g^{\mbox{\tiny\it V}} \, - \, 
g^{\mbox{\tiny\it A}} \, \gamma ^{5} \right ) \psi_{\mbox{\tiny\it n}}$. 
The Green functions $G^{e}_{\pm}$ are essentially the same as 
$G_{\pm \mbox{\tiny\it LL}}$ in Eq.(\ref{eq:eqgpm}) with the
self-evident substitution of $n_{\nu}$ with the electron distribution function
$f_{e}$. Following the same procedure as shown above and employing the 
detailed balance relation satisfied by nucleons in thermal
equilibrium, 
\begin{equation}
S^{\nu\mu}_{np}(- q) \ = \ e^{-\beta ( q^{0} \, + \, 
\Delta \mu _{np}  )} \ S^{\mu\nu}_{pn}(q) \quad ,
\end{equation}
with the difference of the chemical potentials, 
$\Delta \mu _{np} = \mu_{n} - \mu_{p}$, we obtain the collision term 
for the emission and absorption of neutrinos on nucleons as
\begin{equation}
\int \!\! \frac{d^{3}p_{e}}{(2 \pi)^{3}} \ \frac{1}{2E_{e}} \,
\frac{1}{2} \left \{ e^{-\beta ( q^{0} \, + \, 
\Delta \mu _{np}  )} \, f_{e}(p_{e}) \, \left [
\, 1 \, - \, n^{\nu_{e}}(k) \, \right ] \ - \ n^{\nu_{e}}(k) \, 
\left [ \, 1 \, - \, f_{e}(p_{e}) \, \right ] \right \}
\frac{G_{F}^{2}}{2} \, L_{\mu\nu}(k, \, p_{e}) \, S^{\mu\nu}_{pn}(q)
\quad, 
\end{equation}
in the adiabatic mixing case. Here $p_{e}$ and $E_{e}$ are the
momentum and energy of electrons, respectively, and the transfer four
momentum is $q = k - p_{e}$.  Since there is no other components of
self-energy in flavor space than the $\nu_{e}\nu_{e}$ component, the
resulting term is identical to those with no neutrino mixing.

\section{summary}

With a view of application to the simulations of supernova
explosion and proto neutron star cooling, we have derived a Boltzmann 
equation with the neutrino flavor mixing being taken into account. 
The derivation is based on the nonequilibrium field theory, 
and the ordinary gradient
expansion has been performed. We assumed that the typical neutrino 
wave length is much shorter than the scale height of the background
matter distribution, which is true for the supernova cores and proto
neutron stars. The neutrino distribution matrix which is 
non-diagonal in the neutrino flavor space is introduced. 
Following the common practice, the advection part has been obtained 
in the mean field approximation, where the self-energy of neutrino is
non-diagonal in the flavor space. This self-energy gives rise to the
term in the advection part, which is responsible for the neutrino
mixing and does not appear in the ordinary transport equation. 
The collision terms, on the other hand,  have been
calculated in the Born approximation. The collision terms also have 
corrections due to the mixing. In these derivations, 
the relativistic kinematics is taken into consideration. We have
further simplified the Boltzmann equation for the adiabatic flavor mixing, which 
is a good approximation in the supernova cores and proto neutron stars. 
The advection terms thus derived are essentially the same as those
derived by Sirera and P\'{e}rez\cite{sp99}, although they employed the Wigner
function formalism in the mean field approximation and did not give
collision terms. The collision terms derived here, on the other hand, 
have the same structure as those found by Raffelt et al.\cite{rs93a,rs93b} in 
the non-relativistic density matrix method. We have also
shown the general relativistic correction term which accounts for the
red shift and ray bending in the gravitational field and is commonly
taken into account in the supernova and proto neutron star
simulations. 

The applications of the Boltzmann equation found here remain to be
done. Since the corrections due to the flavor mixing are rather
minor, particularly in the case of the adiabatic mixing, it will be 
simple to implement them in the neutrino transport code we have now at 
our disposal\cite{sy99}. This is already underway. Since the mixing angle in
matter is dependent on the neutrino energy and the direction of
momentum with respect to the magnetic field if it exists. In the
analyses of the neutrino flavor mixing in the supernova core, it is
usually assumed that the neutrinos are flowing out
radially\cite{ks96,jr99}. However,
they have an angular distribution near the neutrino sphere. Different 
positions of the resonant conversion due to different directions 
of flight of neutrinos will lead to the reduction of the neutrino
flavor conversion. This will also be true in the absence of the
magnetic field if the energy distribution of neutrinos and the
coupling between neutrinos with different energies are taken into
account. These possibilities and their implications to the mechanism
of the supernova explosion, kick velocity of pulsars, and  
nucleosynthesis of heavy elements will be studied in the forthcoming papers.    

\acknowledgments

I gratefully acknowledge some comments by A. P\'{e}rez. 
This work is partially supported by the 
Grants-in-Aid for the Center-of-Excellence (COE) Research of
the Ministry of Education, Science, Sports and Culture of Japan to
RESCEU (No.07CE2002).


\newpage

\begin{references}
\bibitem{bc89} J. N. Bahcall, Neutrino Astrophysics, Cambridge
               University Press, Cambridge, 1989. 
\bibitem{gr96} G. Raffelt, Stars as Laboratories for Fundamental
               Physics, University of Chicago Press, Chicago, 1996. 
\bibitem{br85}S. W. Bruenn, Astrophys. J. Suppl. {\bf 58}, 771 (1985). 
\bibitem{su94} H. Suzuki, Physics and Astrophysics of Neutrinos, 
               edited by M. Fukugita and A. Suzuki, 
               Springer-Verlag, Tokyo, 1994, p763.
\bibitem{lq66} R. W. Lindquist, Annals of Physics. {\bf 37}, 487 (1966).
\bibitem{mm89} A. Mezzacappa, and R. A. Matzner, Astrophys. J. 
              {\bf 343}, 853 (1989).
\bibitem{sy99} S. Yamada, H. -Th. Janka, and H. Suzuki,
               Astron. \& Astrophys. {\bf 344}, 533 (1999).
\bibitem{fy94} M. Fukugita and T. Yanagida, Physics and Astrophysics of Neutrinos, 
               edited by M. Fukugita and A. Suzuki, 
               Springer-Verlag, Tokyo, 1994, p1.
\bibitem{ru90} M. A. Rudzsky, Astrophysics and Space Sciences 
              {\bf 165}, 65 (1990).
\bibitem{rs93a} G. Raffelt, G. Sigl and L. Stodolsky, Phys. Rev. Lett. 
               {\bf 70}, 2363 (1993).
\bibitem{rs93b} G. Raffelt and G. Sigl, Astropart. Phys. {\bf 1}, 165 (1993).
\bibitem{pa95} J. Pantaleone, Phys. Lett. {\bf B342}, 250, (1995). 
\bibitem{dn96} J. C. D'Olivo and J. F. Nieves, Int. J. Mod. Phys. 
              {\bf A11}, 141 (1996).
\bibitem{sp99} M. Sirera and A. P\'{e}rez, Phys. Rev. {\bf D59}, 125011 
              (1999).
\bibitem{lo95} F. N. Loreti, Y. -Z. Qian, G. M. Fuller and
               A. B. Balantekin, Phys. Rev. {\bf D52}, 6664 (1995).
\bibitem{be97} F. Buccella, S. Esposito, C. Gualdi and G. Miel, 
               Z. Phys. {\bf C73}, 633 (1997).  
\bibitem{ks96} A. Kusenko and G. Segre, Phys. Rev. Lett. {\bf 77},
               4872 (1996). 
\bibitem{jr99} H. -Th. Janka and G. Raffelt, Phys. Rev. {\bf D59},
               023005 (1999). 
\bibitem{sw61} J. Schwinger, J. Math. Phys. {\bf 2}, 407 (1961). 
\bibitem{kd65} L. V. Keldysh, JETP {\bf 20} 1018 (1965).
\bibitem{ch85} K. -C. Chou, Z. -B. Su, B. -L. Hao and L. Yu, 
               Phys. Rep. {\bf 118}, 1 (1985).
\bibitem{dz84} P. Danielewicz, Annals of Physics {\bf 152}, 239 (1984).
\bibitem{mh88} P. D. Mannheim, Phys. Rev. {\bf D37}, 1935 (1988).
\bibitem{nr88} D. N\"{o}tzold and G. Raffelt, Nucl. Phys. {\bf B307},
               924 (1988).
\bibitem{ec96} S. Esposito and G. Capone, Z. Phys. {\bf C70}, 55 (1996).
\bibitem{eg96} P. Elmfors, D. Grasso and G. Raffelt, Nucl. Phys. 
              {\bf B479}, 3 (1996). 
\bibitem{kb62} L. P. Kadanoff and G. Baym, Quantum Statistical
               Mechanics, The Benjamin/Cummings Publishing Company,
               1962, London. 
\bibitem{sc90} P. J. Schinder, Astrophys. J. Suppl. {\bf 74}, 249 (1990).
\bibitem{yt99} S. Yamada and H. Toki, Phys. Rev. {\bf C61}, 015803 
              (2000).
\end{references}
\end{document}